%% file: ANT.tex
\documentclass[10pt,journal]{IEEEtran}

\usepackage{cite}
\usepackage[OT1]{fontenc} 
\usepackage{hyperref}       
\usepackage{url}            
\usepackage{booktabs}       
\usepackage{amsfonts}       
\usepackage{nicefrac}       
\usepackage{microtype}      
\usepackage{lipsum}		
\usepackage{amsmath}
\usepackage[english]{babel}
\usepackage{algorithm}
\usepackage{algpseudocode}
\usepackage{float}
\usepackage{multirow}
\usepackage[table,xcdraw]{xcolor}
\usepackage{pgfplots}
\usepackage{tikz}           
\usetikzlibrary{matrix}
\usepgfplotslibrary{groupplots}
\pgfplotsset{compat=newest}

\usepackage{pgfplots}
\usepgfplotslibrary{groupplots}
\usetikzlibrary{pgfplots.groupplots}
\usetikzlibrary{plotmarks}
\usetikzlibrary{calc}
\usepgfplotslibrary{external}
\usepackage[font=small,labelfont=bf]{caption}
\usepackage{subcaption}

\definecolor{airforceblue}{rgb}{0.36, 0.54, 0.66}
\definecolor{aliceblue}{rgb}{0.94, 0.97, 1.0}

\definecolor{cadmiumgreen}{rgb}{0.0, 0.42, 0.24}
\definecolor{honeydew}{rgb}{0.94, 1.0, 0.94}

\definecolor{carnelian}{rgb}{0.7, 0.11, 0.11}
\definecolor{melon}{rgb}{0.99, 0.74, 0.71}

\definecolor{orange-red}{rgb}{1.0, 0.27, 0.0}
\definecolor{peach-orange}{rgb}{1.0, 0.8, 0.6}

\definecolor{pastelyellow}{rgb}{1.0, 0.92, 0.71} 
\definecolor{uscgold}{rgb}{1.0, 0.8, 0.0}

\definecolor{purpleheart}{rgb}{0.41, 0.21, 0.61}
\definecolor{lightmauve}{rgb}{0.86, 0.82, 1.0}

\definecolor{darkbrown}{rgb}{0.4, 0.26, 0.13}
\definecolor{moccasin}{rgb}{0.98, 0.92, 0.84}

\definecolor{ashgrey}{rgb}{0.38, 0.38, 0.38}
\definecolor{lightgray}{rgb}{0.90, 0.90, 0.90} 
\definecolor{cadetgrey}{rgb}{0.57, 0.64, 0.69}

\definecolor{black}{rgb}{0.0, 0.0, 0.0}
\definecolor{white}{rgb}{1.0, 1.0, 1.0}

\usepackage{graphics}

\ifCLASSINFOpdf
\else
\fi
\begin{document}
%
\title{ Adversarial Network Traffic: \\Towards Evaluating the Robustness of Deep Learning-Based Network Traffic Classification}
%
%
%

\author{Amir~Mahdi~Sadeghzadeh, 
        Saeed~Shiravi, and Rasool~Jalili
\thanks{A. Sadeghzadeh, S. Shiravi, and R. Jalili are with the Department of Computer Enginnering, Sharif University of Technology, Tehran 11365-11155, Iran, (e-mail: amsadeghzadeh@ce.sharif.edu; saeedshiravi@ce.sharif.edu; jalili@sharif.edu)}}

%
%


\markboth{IEEE Transactions on Network and Service Management,~Vol.~*, No.~*, MONTH~YEAR}%
\markboth{}

%



\maketitle

\begin{abstract}
 Network traffic classification is used in various applications such as network traffic management, policy enforcement, and intrusion detection systems. 
 Although most applications encrypt their network traffic and some of them dynamically change their port numbers, Machine Learning (ML) and especially Deep Learning (DL)-based classifiers have shown impressive performance in network traffic classification. In this paper, we evaluate the robustness of DL-based network traffic classifiers against Adversarial Network Traffic (ANT).
 ANT causes DL-based network traffic classifiers to predict incorrectly using Universal Adversarial Perturbation (UAP) generating methods. Since there is no need to buffer network traffic before sending ANT, it is generated live. We partition the input space of the DL-based network traffic classification into three categories: packet classification, flow content classification, and flow time series classification. To generate ANT, we propose three new attacks injecting UAP into network traffic.
AdvPad attack injects a UAP into the content of packets to evaluate the robustness of packet classifiers. AdvPay attack injects a  UAP into the payload of a dummy packet to evaluate the robustness of flow content classifiers. AdvBurst attack injects a specific number of dummy packets with crafted statistical features based on a UAP into a selected burst of a flow to evaluate the robustness of flow time series classifiers. 
 The results indicate injecting a little UAP into network traffic, highly decreases the performance of DL-based network traffic classifiers in all categories.
\end{abstract}


%
\IEEEpeerreviewmaketitle

\newcommand{\comment}[1]{}

\section{Introduction}
%
%
%
%

\label{Int}
\input{content/introduction}

\section{Background}
\label{sec:Background}
\input{content/background}

\section{Related Works}
\label{sec:Related Works}
\input{content/related}

\section{Adversarial Network Traffic}
\label{sec:Adversarial Network Traffic}
\input{content/proposed_approach}

\section{Evaluation}
\label{sec:Evaluation}
\input{content/evaluation}

\section{Results Analysis}
\label{sec:resana}
\input{content/resultanalysis}

\section{ANT In Practice}
\label{sec:discussion}

\input{content/discussion}

\section{Conclusion}
\label{sec:Conclusion}
\input{content/conclusion}

\section*{Acknowledgment}

\input{content/Acknowledgement}
\bibliographystyle{IEEEtran}
\bibliography{ref}
\begin{IEEEbiography}[{\includegraphics[width=1in,height=1.25in,clip,keepaspectratio]{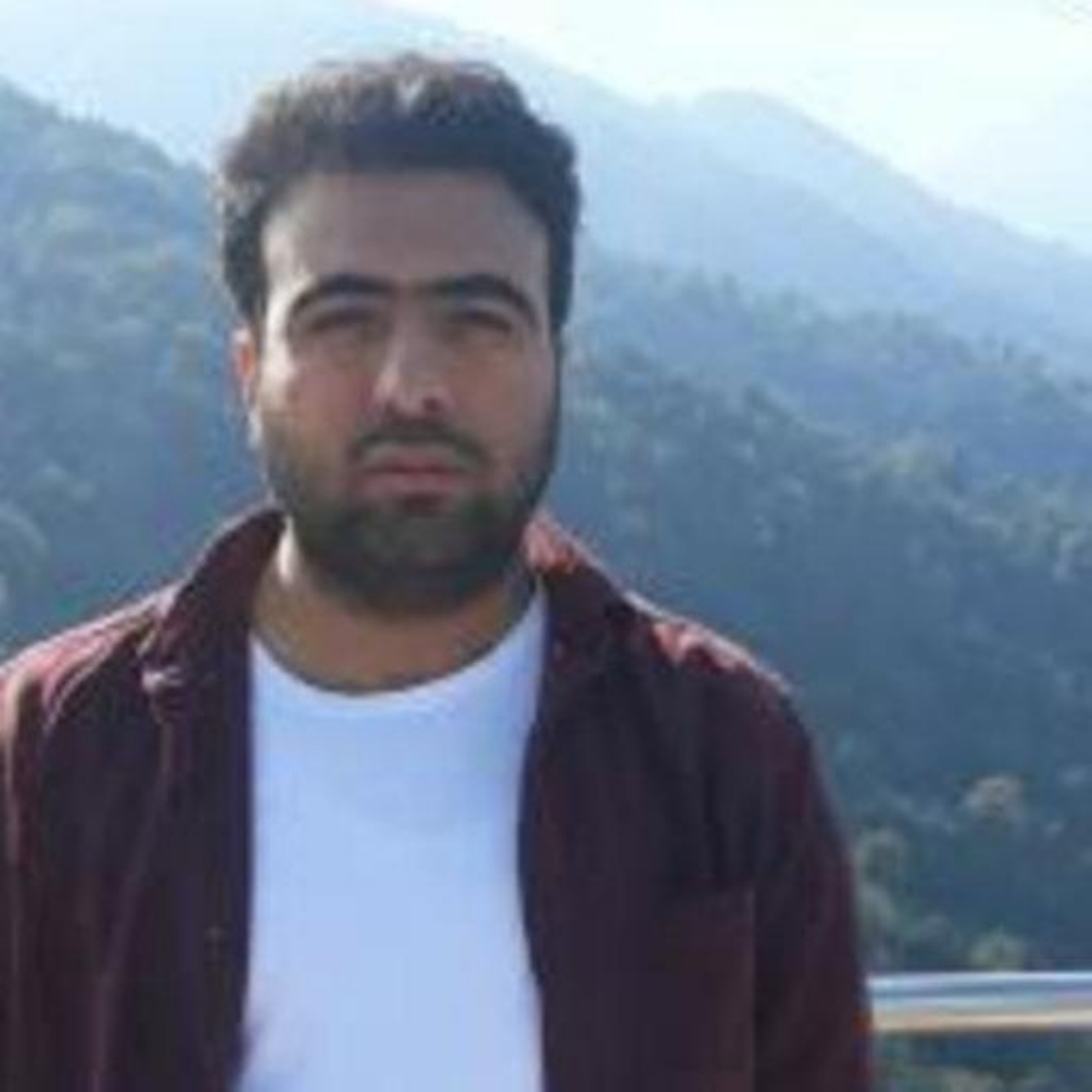}}]{Amir Mahdi Sadeghzadeh}
	received his B.Sc. degree in Information Technology Engineering from Isfahan University of Technology in 2014 and his M.Sc. degree in Information Technology Engineering from Sharif University of Technology in 2016. He is currently a Ph.D. candidate at the Department of Computer Engineering, Sharif University of Technology. His research interests include Deep Learning Security, Adversarial Deep Learning, and Network traffic Classification.
\end{IEEEbiography}
\vspace{-1cm}
\begin{IEEEbiography}[{\includegraphics[width=1in,height=1.25in,clip,keepaspectratio]{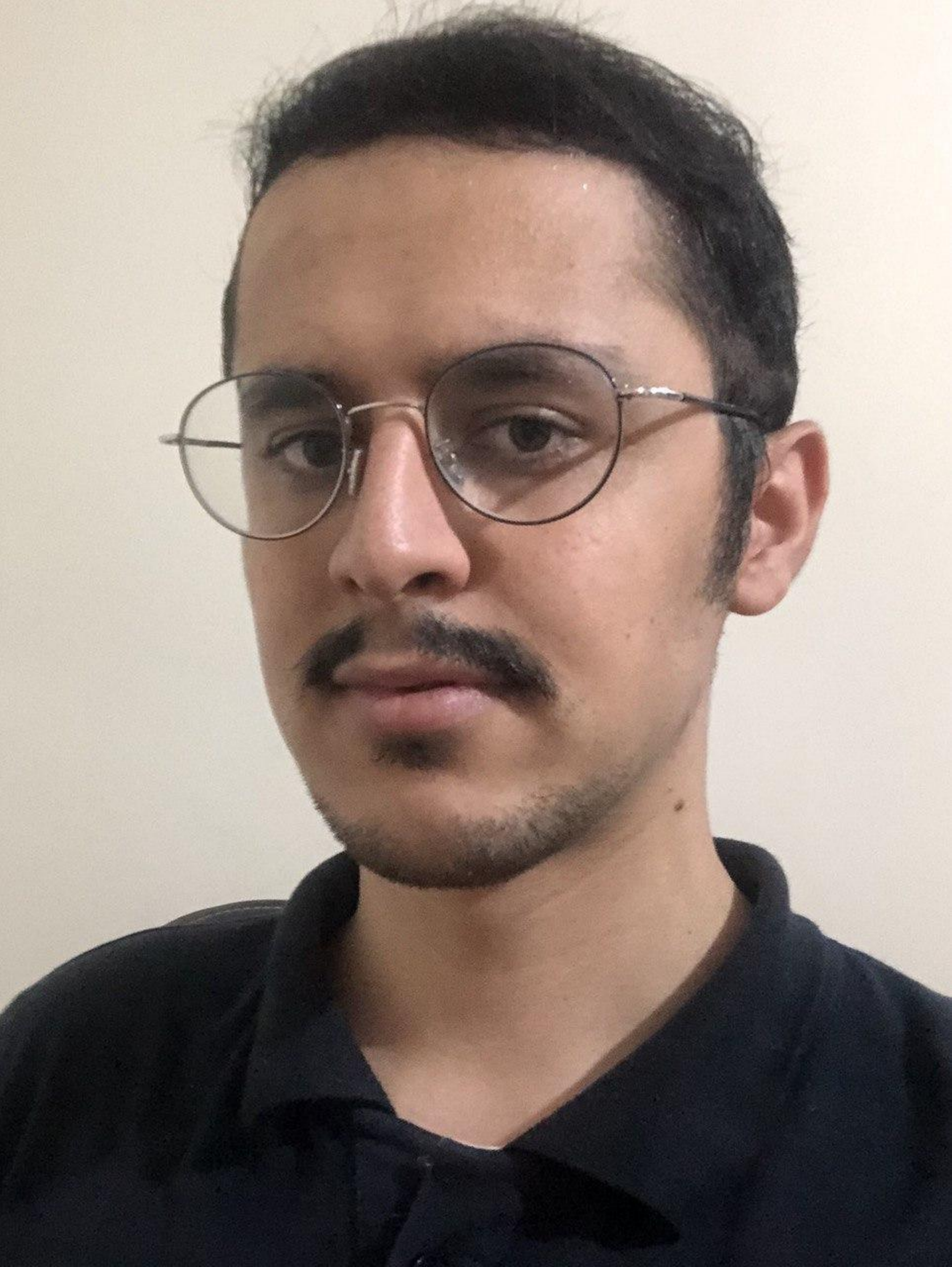}}]{Saeed Shiravi}
	received his B.Sc degree in Computer Engineering from University of Isfahan in 2015 and M.Sc. degree in Information Technology Engineering from Sharif University of Technology in 2018.  His research interests include Anonymity Network, Network Traffic Classification, and Adversarial Machine Learning.
\end{IEEEbiography}
\vspace{-1cm}
\begin{IEEEbiography}[{\includegraphics[width=1in,height=1.25in,clip,keepaspectratio]{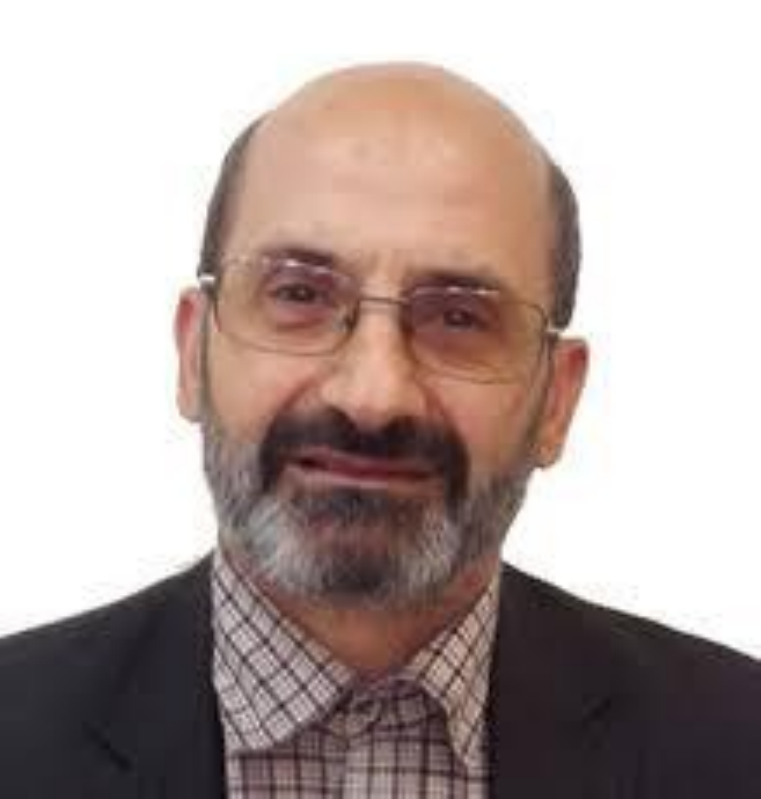}}]{Rasool Jalili}
	received his B.Sc. degree in Computer Science from Ferdowsi University of Mashhad in 1985, and his M.Sc. in Computer Engineering from Sharif University of Technology in 1989. He received his Ph.D. in Computer Science from The University of Sydney, Australia, in 1995. He then joined the Department of Computer Engineering, Sharif University of Technology, Tehran, Iran, in 1995. He has published more than 150 papers in Computer Security and Pervasive Computing in international journals and conferences proceedings. He is now an associate professor and the director of Data and Network Security Lab (DNSL) in Sharif University of Technology. His research interests include Access Control, Vulnerability Analysis, Database Security, and Machine Learning Security.
\end{IEEEbiography}

\end{document}

%% file: content/introduction.tex
\IEEEPARstart{IN}{} recent years, Internet traffic has grown due to the emergence of new applications and services in market. Hence, management of network traffic is more challenging than ever and the necessity of network traffic classification is obvious in many cases, such as quality of service provision, resource usage management, billing in ISPs, anomaly detection, and Policy Enforcement systems. So far, three main approaches to network traffic classification have been introduced, including port-based, payload-based, and ML-based classification\cite{DBLP:journals/comsur/PachecoEGBA19}. 
\comment{
In the early stage of Internet, most applications used well-known port numbers, and most traffic classification tasks were solved by extracting port numbers from TCP/UDP headers and assigning them to the predefined list introduced by 
IANA \cite{IANA}.
 Although port-based approach has prevailed in many devices,
 an adversary can use random port numbers or well-known port numbers to evade port-based classifiers\cite{DBLP:journals/cn/DusiGS11}. 
Payload-based approach 
searches for specific signatures in network traffic and matches predefined patterns to the extracted content of packets. 
Although this approach has higher accuracy and robustness in comparison with port-based approaches, 
an adversary can encrypt his/her network traffic or obfuscate the pattern in the content of packets to evade payload-based classifiers\cite{DBLP:journals/comsur/PachecoEGBA19,DBLP:journals/cn/DusiGS11}.
}
Although port-based and payload-based approaches have prevailed in many devices, an adversary can use random port numbers or well-known port numbers to evade port-based classifiers \cite{DBLP:journals/cn/DusiGS11}, and encrypt his/her network traffic or obfuscate the pattern in the content of packets to evade payload-based classifiers\cite{DBLP:journals/comsur/PachecoEGBA19,DBLP:journals/cn/DusiGS11}.

ML-based approaches utilize ML algorithms for network traffic classification. The byte sequence and statistical features of packets or flows are the most common features that are used by ML-based network traffic classifiers. 
Although ML-based approaches have been put forward with claims of fixing the problems of previous approaches, the robustness of ML-based classifiers has not been evaluated. In this paper, we will investigate the robustness of such classifiers in a comprehensive way.
In recent years, Deep Neural Networks (DNNs), also called Deep Learning (DL), have shown great success in solving complex problems such as image classification \cite{DBLP:conf/nips/KrizhevskySH12}, and speech recognition \cite{6296526}. 
 These advances have motivated the researchers to use DNNs in other domains, and recently, researchers have demonstrated DNNs have promising results in network traffic classification \cite{DBLP:journals/access/MartinCSL17,DBLP:journals/cn/Caicedo-MunozEC18,DBLP:journals/tnsm/AcetoCMP19,DBLP:conf/ccs/SirinamIJW18,DBLP:conf/ndss/RimmerPJGJ18,DBLP:journals/soco/LotfollahiSZS20}.
In this paper, we focus on DL-based network traffic classifiers, which are state of the art in ML-based network traffic classification.

Based on previous studies, we divide the input space of DL-based network traffic classification into three categories: Packet Classification (PC), Flow Content Classification (FCC), and Flow Time Series Classification (FTSC). In PC, the byte sequence of a packet is given to a classifier, and each packet is labeled. In FCC, the byte sequence of the first n packets of a flow is fed to a classifier, and the class of each flow is predicted. In FTSC, the sequence of statistical features of the first m packets of a flow, such as packets size and inter-arrival times between packets, are passed to a classifier, and each flow is classified. 
We consider two classifiers in each input space category to assess the influence of various items in the input of classifiers.
In this paper, we use One-Dimensional Convolutional Neural Networks (1D-CNN) to classify ISCXVPN2016 dataset \cite{DBLP:conf/icissp/Draper-GilLMG16} in which multiple application, including Skype, Facebook, and Hangouts generate network traffic with various types. The target of network traffic classification in this study is traffic characterization in which the type of network traffic such as VoIP, streaming, email, and chat is determined.

After the success of DNNs in various classification tasks, their robustness has become the subject of much debate \cite{DBLP:journals/corr/SzegedyZSBEGF13,DBLP:conf/sp/Carlini017,DBLP:journals/corr/GoodfellowSS14,DBLP:journals/corr/abs-1712-09665}. In 2014, Szegedy \textit{et al.} \cite{DBLP:journals/corr/SzegedyZSBEGF13} have demonstrated that DNNs are vulnerable to adversarial examples, which are maliciously crafted inputs that cause a classifier to make a mistake. 
This vulnerability questioned the robustness of DL-based classifiers in adversarial environment in which an adversary can interfere in the training and the prediction phases. In this paper, the robustness of DL-based network traffic classifiers is challenged against Adversarial Network Traffic (ANT). ANT is adversarially crafted network traffic in which the content or the statistical features of network traffic is perturbed. 

There are two serious constraints to generate ANT.
First, in some types, such as VoIP, email, and chat, the content and the statistical features of a flow are generated over time and are dependent to the user, and there is no access to the entire flow at the beginning of it. Second, a flow is generated by two entities, and neither of them knows the content and the statistical features of entire flow to make adversarial perturbation based on it.
According to these constraints, we use Universal Adversarial Perturbation (UAP) generating methods to generate ANT. Moosavi-Dezfooli \textit{et al.} \cite{DBLP:conf/cvpr/Moosavi-Dezfooli17} demonstrate that there is a perturbation that, if it is added to a set of data, it will cause DNNs to make incorrect predictions on most of them, and this perturbation is not unique. In ANT, UAP is generated on a pre-collected set of network traffic, and whenever a new flow or a new packet is being sent, the pre-made UAP is injected into it. This approach does not need to have access to the entire data to make perturbation, and UAP is made beforehand on a pre-collected set of network traffic. Consequently, ANT can be generated live, and there is no need to buffer target network traffic.

In this paper, three new attacks for generating ANT are proposed, including Adversarial Pad (AdvPad) attack, Adversarial Payload (AdvPay) attack, and Adversarial Burst (AdvBurst) attack. 
AdvPad injects a UAP into the end or the start of packets payload to evaluate the robustness of packet classifiers. AdvPay injects a UAP into the payload of a dummy packet at the first n packets of a flow to evaluate the robustness of flow content classifiers. AdvBurst injects a specific number of dummy packets with crafted statistical features based on a UAP into the end of a selected burst of a flow to evaluate the robustness of flow time series classifiers.
 Experiments show that all classifiers are vulnerable to ANT, and by injecting a little UAP into network traffic, the performance of DL-based network traffic classifiers highly decreases. Results demonstrate that the researchers should investigate the methods for improving the robustness of DL-based network traffic classifiers.

The main contributions of this paper are as follows:
\begin{itemize}
	\item 
	Network traffic is formally defined, including packets, unidirectional flows, bidirectional flows, and three input space categories of DL-based network traffic classifiers. 
	\item 
	The influence of input space categories on the performance and the robustness of DL-based network traffic classifiers is investigated.
	\item 
	Adversarial Network Traffic (ANT) is proposed to evaluate the robustness of DL-based network traffic classifiers. 
	\item 
	Three new attacks are proposed to generate ANT, including AdvPad, AdvPay, and AdvBurst attacks.
\end{itemize}

The rest of the paper is organized as follows. In Sec. \ref{sec:Background}, network traffic, and three input space categories of DL-based network traffic classification are formally defined. Also, the DNNs and basic methods for generating adversarial examples are introduced. Sec. \ref{sec:Related Works} reviews related works on  DL-based network traffic classification. In Sec. \ref{sec:Adversarial Network Traffic}, ANT will be presented, and AdvPad, AdvPay, and AdvBurst attacks are proposed. Sec. \ref{sec:Evaluation} evaluates the robustness of classifiers in three input space categories against three proposed attacks. 
Sec. \ref{sec:resana} analyses the results of experiments.
Sec. \ref{sec:discussion} discusses how to use ANT in practice.
Finally, in Sec. \ref{sec:Conclusion}, the conclusions of this work will be discussed.

%% file: content/background.tex
Network traffic, DL-based classifiers, and adversarial examples are the fundamental constituents of the DL-based network traffic classification and ANT. Accordingly, we introduce these ingredients in the following subsections.
\subsection{Formal Specification of Network Traffic}
Network traffic consists of bidirectional flows and packets, which are the common objects for network traffic classification. Each packet consists of multiple headers and a payload in which application data exists. 
We indicate the headers of a packet by 5-tuple information, which stems from the content of the IP  layer and Transport Layer (TL) headers.

\textbf{Definition 1}. An IP packet $P_i$ is a sequence of an IP layer header, a TL header, and a payload.
\footnotesize
\begin{equation}
\begin{split}
	&P_i = \: <H^{IP}_{p_i},H^{TL}_{p_i},Pay_{p_i}> , \\
&H^{IP}_{P_i} = (IP^{src}_{P_i}, IP^{dst}_{P_i} ), \\
&H^{TL}_{P_i} = (Port^{src}_{P_i}, Port^{dst}_{P_i}, Proto_{p_i}),
\end{split}
\end{equation}
\normalsize
where $IP^{src}_{P_i}$ and $IP^{dst}_{P_i}$ indicate source and destination IP addresses of the packet, respectively. $ Port^{src}_{P_i}$ and $Port^{dst}_{P_i}$ are source and destination ports of the transport layer, respectively, and $Proto_{p_i}$ is the transport layer protocol of the packet.
Unidirectional flow is a set of packets that share the same source and destination information and transport layer protocol. 

\textbf{Definition 2}. A unidirectional flow $UF_i$ is a set of an IP layer header, a TL header, a timeout value,  and a sequence of packets.
\footnotesize
\begin{equation}
\begin{split}
&UF_i = \: \{H^{IP}_{UF_i},H^{TL}_{UF_i},P_{UF_i},TimeOut_{UF_i}\} , \\
&H^{IP}_{UF_i} = (IP^{src}_{UF_i}, IP^{dst}_{UF_i} ), \\
&H^{TL}_{UF_i} = ( Port^{src}_{UF_i}, Port^{dst}_{UF_i}, Proto_{UF_i}), \\
&P_{UF_i} = <P^i_1, P^i_2, ..., P^i_m>, \\
s.t.\;&  H^{IP}_{P^i_j} = H^{IP}_{UF_i}, \;H^{TL}_{P^i_j} = H^{TL}_{UF_i}, j \in [1,m], \\
& IAT_{P^i_{k+1},P^i_k} \leq TimeOut_{UF_i}, k \in [1,m-1],
\end{split}
\end{equation}
\normalsize
where $m$ is the number of packets, $IAT_{P^i_{k+1},P^i_k}$ is inter-arrival time between $k$ and $k+1$ packets of the packet sequence $P_{UF_i}$ in the unidirectional flow $UF_i$. $IP^{src}_{UF_i}$ and $IP^{dst}_{UF_i}$ are source and destination IP addresses of $UF_i$, respectively. $ Port^{src}_{UF_i}$ and $Port^{dst}_{UF_i}$ are source and destination ports of the transport layer, respectively, and $Proto_{UF_i}$ is the transport layer protocol of $UF_i$. A bidirectional flow is the union of two unidirectional flows, where IP addresses and ports have switched.

\textbf{Definition 3}. A bidirectional flow $BF_i$ is the union of two opposite unidirectional flows $UF_j$ and $UF_k$.
\footnotesize
\begin{equation}
\begin{split}
&BF_i = \: UF_j \cup UF_k, \\
s.t.\; &IP^{src}_{UF_j} = IP^{dst}_{UF_k}, \: IP^{dst}_{UF_j} = IP^{src}_{UF_k},\\
& \; Port^{src}_{UF_j} = Port^{dst}_{UF_k}, \: Port^{dst}_{UF_j} = Port^{src}_{UF_k}, \\
& Proto_{UF_j} = Proto_{UF_k},\; TimeOut_{UF_j} = TimeOut_{UF_k}.
\end{split}
\end{equation}
\normalsize
We focus on bidirectional flow in this paper and  use flow instead of bidirectional flow in the rest of the paper for simplicity.

\subsection{ML-Based Network Traffic Classification}
\label{sec:ML_Based_Network_Traffic_Classification}
Classification is a subset of supervised learning that categorizes a set of data into classes. A ML-based classifier is a function $f$ which maps an input space $\mathcal{X}$ to an output space $\mathcal{Y}=\{y_1, ... , y_k\}$ where $y_i$ is $i^{th}$ class, and $k$ is the number of classes.  
In the training phase of a classifier, we need to have a training set in which the true class of each sample has been given $\{(x_i,y_i)\}_{i=1}^{n}$, where $n$ is the number of samples in the training set and $y_i$ is the true class (label) of the sample $x_i$.  A good classifier generalizes to unseen samples in the test set and classifies them correctly. 

According to the previous studies, network traffic classification can be used in several domains, including protocol detection, application identification, website and video fingerprinting, and traffic characterization. 
In this paper, we focus on traffic characterization, which specifies the type of network traffic, such as VoIP, streaming, file transferring, and email. 
Traffic characterization is fundamental in many scenarios, such as network traffic accounting, quality of service providing, and policy enforcement. 

 We partition the input space of network traffic classification into three categories:
($i$) packet classification (PC), ($ii$) flow content classification (FCC), and ($iii$) flow time series classification (FTSC).
In what follows, each input space category will be explained.
Since the information in the header of the IP layer is dependent on the machines that generate network traffic, we do not consider IP header for network traffic classification.

\textbf{Packet Classification (PC)}. 
In packet classification, the byte sequence of a packet is given to a classifier, and it labels each packet separately. Based on the literature, we consider two versions of packet classification. In the first version (PC-HP), TL header and payload, and in the later version (PC-P), only payload of a packet is given to classifier. For packet $P_i$, we have:
\footnotesize
\begin{equation}
\begin{split}
\textrm{PC-HP Input for $P_i$:} &<H^{TL}_{p_i},Pay_{p_i}>,\\
 \textrm{PC-P Input for $P_i$:} &<\!Pay_{p_i}>.
\end{split}
\end{equation}
\normalsize

\textbf{Flow Content Classification (FCC)}.
In this category, the byte sequence of the first n packets of a flow is given to a classifier, and each flow is labeled. 
FCC has two versions. In the first version (FCC-HP), TL headers and payloads, and in the second one (FCC-P), only payloads of the first n packets of a flow are given to a classifier. To improve the performance of classifiers, we multiply the sign of packet direction to the content of that packet. The direction sign is considered positive (+1) for packets from source (e.g., client) to
destination (e.g., server) and negative (-1) for packets from destination to source. For flow $F_i$, we have:
\footnotesize
\begin{equation}
\begin{split}
\label{FCC-input}
\textrm{FCC-HP Input for $F_i$:}& <H^{TL}_{p^i_0} \times D_{p^i_0},Pay_{p^i_0} \times D_{p^i_0},\\
& ..., H^{TL}_{p^i_{n-1}} \times D_{p^i_{n-1}},Pay_{p^i_{n-1}}\times D_{p^i_{n-1}}>, \\
\textrm{FCC-P Input for $F_i$:}& <Pay_{p^i_0}\times D_{p^i_0}, ..., ,Pay_{p^i_{n-1}}\times D_{p^i_{n-1}}> ,
\end{split}
\end{equation}
\normalsize
where $D_{p^i_j} \in \{+1,-1\}$ is $j^{th}$ packet direction of flow $F_i$.

\textbf{Flow Time Series Classification (FTSC)}.
In this category, the statistical features of the first m packets of a flow are given to a classifier.
Two versions of FTSC have been considered in the previous studies. In the first version (FTSC-IAT), inter-arrival times between packets of a flow, and in the second version (FTSC-PS), packets sizes of a flow are given to a classifier in time-series format. Similar to FCC, we multiply the sign of packets direction to the statistical features of packets to improve the performance of the classifiers. For flow $F_i$, we have:
\footnotesize
\begin{equation}
\label{eq:FTSC_data}
\begin{split}
\textrm{FTSC-IAT Input for $F_i$:} &<IAT_{(p^i_0,p^i_1)} \times D_{p^i_1}, IAT_{(p^i_1,p^i_2)} \\
\times D_{p^i_2}, &..., IAT_{(p^i_{m-2},p^i_{m-1})} \times D_{p^i_{m-1}}>, \\
\textrm{FTSC-PS Input for $F_i$:} &<PS_{p^i_0}\times D_{p^i_0}, PS_{p^i_1}\times D_{p^i_1},\\
& ..., PS_{p^i_{m-1}}\times D_{p^i_{m-1}}>,
\end{split}
\end{equation}
\normalsize
where $IAT_{(p^i_j,p^i_k)} $ is  inter-arrival time between $j^{th}$ and $k^{th}$ packets, 
and $PS_{p^i_j}$ is the size of $j^{th}$ packet in flow $F_i$.

\subsection{Deep Neural Networks}
 Deep neural networks (DNNs) are a function $y_i = f_{DNN}(x_i)$ which takes an input  $x_i \in \mathcal{X}$ and returns an output $y_i \in \mathcal{Y}$. 
 DNNs act on raw data and do not need feature engineering. Therefore, there is no need to have feature selection or an expert who extracts the most salient features of network traffic.
Convolutional Neural Network (CNN), Recurrent Neural Network (RNN), and Stacked Denoising Autoencoders (SDAE) are the three main DNNs which have been used in network traffic classification. Previous investigations \cite{DBLP:journals/tnsm/AcetoCMP19,DBLP:conf/ccs/SirinamIJW18} have demonstrated that the One-Dimensional Convolutional Neural Networks (1D-CNNs) have delivered the best performance in network traffic classification. In this paper, we use 1D-CNN to classify network traffic. 


Each neural network includes an input layer, multiple hidden layers, and an output layer. Each layer consists of multiple neurons that are connected to neurons in the adjacent layers. 
 Except for the input layer, the output of a given layer is calculated through applying a nonlinear function on an obtained value from multiplying the output of previous layer by the weights matrix of that layer. \comment{For  a deep neural network $f_{DNN}$ with $l$ layers and an input $x$, we have:
\small
\begin{equation}
\begin{split}
&f^i_{DNN}(x^i) = \sigma(\theta^i \times x^i + b^i), \\
&f_{DNN}(x) = f^{l-1}_{DNN}(f^{l-2}_{DNN}(..., f^1_{DNN}(x))),
\end{split}
\end{equation}
\normalsize
where, $f^i_{DNN}(x)$ is the output of $i^{th}$ layer, $x^i$ is the input of the $i^{th}$ layer, $\sigma$ is a non-linear function, $\theta^i$ is the weights matrix of $i^{th}$ layer, and $b_i$ is the bias vector of $i^{th}$ layer.} 
A DNN assigns the label \footnotesize$\hat{y} = \underset{0\leq i<k}{\mathrm{argmax}}\,f_{DNN}(x)_i$ \normalsize to the input $x$, where $f_{DNN}(x)_i$ is the $i^{th}$ element of DNN output. 
A convolutional layer uses filters that convolve on the input of the layer to extract local features. A deep convolutional neural network consists of multiple convolutional layers at the start of the network to extract features of the input and a fully connected neural network that uses extracted features by the convolutional layers to classify input data.
\comment{
 Given a one-dimensional input of size N and a one-dimensional filter of size M, the $n^{th}$ element of the $i^{th}$ layer output $f^{i,n}_{DCNN}$ is defined as follows:

\small
\begin{equation}
f^{i,n}_{DCNN}(x) = \sum_{m=0}^{M-1}\sigma(\theta^i_m \times x_{n+m} + b^i_m), \; s.t. \; 0 \leq n \leq N-M
\end{equation}
\normalsize
where, $\theta^i_m$ is the $m^{th}$ element of $i^{th}$ layer's filter weights, $x_{n+m}$ is the $(n+m)^{th}$ element of $i^{th}$ layer input, and $b^i_m$ is the $m^{th}$ element of $i^{th}$ layer filter bias. 
}
In the training phase, weights matrix of all layers $\theta$
are adjusted to minimize a loss function $J$ which measures the distance between predicted label $\hat{y}$ and the true label $y$. 
Hence, training is a minimization problem on the weights of classifier. The output of the training phase is $\theta^{*}$ that minimizes the loss function $J$. 
\comment{We have:
\small
\begin{equation}
\theta^{*} = \underset{\theta}{\mathrm{argmin}} \; J(\theta, x, y)
\end{equation}
\normalsize
}
 Stochastic gradient descent‌ (SGD) and its versions are used for the minimization of loss function in DNNs. 
 \comment{
\small
\begin{equation}
\theta^{t+1} = \theta^{t} - \eta \nabla_{\theta^t} J(\theta^{t}, x, y)
\end{equation}
\normalsize
where, $\theta^{t}$ is weights of DNN at $t^{th}$ iteration, $\eta$ is  learning rate which tunes the size of change in weights, and $\nabla_{\theta^t} J(\theta,x, y)$ is the gradient of loss function $J$ with respect to $\theta^t$.
}
Previous studies have used batch normalization, dropout, and polling layers to improve the performance of DNNs in network traffic classification. 
(For more information, see \cite{Goodfellow-et-al-2016}).

\subsection{Adversarial Example}
\label{BG_AE}
Although DNNs have achieved great success in solving complex problems, they demonstrate serious vulnerability against adversarial examples \cite{DBLP:journals/corr/SzegedyZSBEGF13}. An adversarial example is a crafted input which causes the target classifier to make a mistake.   Suppose that a classifier $f^*$ assigns the true label to each data and the target classifier of adversarial example attack is $f$. For an adversarial example $x'$, we have:
\footnotesize
\begin{equation} 
\begin{split}
&f^*(x') = y,\\
& f(x') = y',\quad s.t.\: y'\, \neq y.\\
\end{split}
\end{equation}
\normalsize
The main method for creating an adversarial example is adding an adversarial perturbation $\xi$ to a real data $x$. In this method, it is supposed that after adding perturbation to data x, the true class of $x' = x + \xi$ and $x$ must be the same. Given that a large perturbation can change the true class of data, the amount of perturbation has to be limited in many contexts such as, image classification.
The most common distant function is used in the literature is  $L_p = \left\lVert x'-x\right\rVert_p$ so that $P$-norm $\left\lVert . \right\rVert_p$ is defined as:
\footnotesize
\begin{equation}
\left\lVert d \right\rVert_p =\biggl (\sum_{i=0}^{D-1}|d_i|^p\biggr )^{1/p}
\end{equation}
\normalsize
where $D$ is the dimension of data $d$ and $p \in [0,\infty]$.
Generation of an adversarial example can be formulated as a minimization problem: 
\footnotesize
\begin{equation}
\begin{split}
\label{AE_MP}
&\underset{x'}{\mathrm{arg \,min}}  \;\left\lVert x' - x \right\rVert_p\\
s.t.\:  f^*(x') = y, \;f(x') &= y', \;y'\, \neq y, \;\;x' \in Domain(x).
\end{split}
\end{equation}
\normalsize
Goodfellow \textit{et al.} \cite{DBLP:journals/corr/GoodfellowSS14} propose Fast Gradient Sign Method (FGSM) attack to solve the minimization problem. The authors use the gradient of DNN loss function to compute adversarial perturbation, and $L_{\infty}$ to limit the size of perturbation. 
In FGSM, adversarial perturbation is calculated as:
\footnotesize
\begin{equation}
\xi = \epsilon. sign(\nabla_x J(\theta, x, y))
\end{equation}
\normalsize
where $\nabla_x J(\theta, x, y)$ is the gradient of DNN loss function with respect to input $x$. 
\comment{Kurakin \textit{et al.} \cite{DBLP:conf/iclr/KurakinGB17a} propose an iterative gradient sign (IGS) attack in which instead of taking one step with size $\epsilon$, they take multiple steps with size $\alpha$, which is smaller than $\epsilon$. 
	They applied $clip$ function to limit the maximum change of each dimension to $\epsilon$.
In the first step, $x_0'=x$, and in each iteration:
\small
\begin{equation}
\begin{split}
&\xi_t = \alpha. sign(\nabla_x J(\theta, x_t', y)),\\
&x_{t+1}' = clip_\epsilon (x_t ' + \xi_t).
\end{split}
\end{equation}
\normalsize
Authors demonstrate that adversarial examples crafted by IGS outperform FGSM in terms of the size of perturbation. 
}

Moosavi-Dezfooli \textit{et al.} \cite{DBLP:conf/cvpr/Moosavi-Dezfooli17} introduce the Universal Adversarial Perturbation (UAP) attack, which is different from previous attacks. In previous attacks, a perturbation is made for each data. However, the UAP attack adds a single perturbation $\xi^u$ to a set of data and hops that they would be misclassified with a high probability.
\comment{This attack has two constraints:
\small
\begin{equation}
\begin{split}
&\left\lVert \xi^u \right\rVert_p < \epsilon\\
&P_{x\in\mathcal{X}}( func(x+\xi^u) \neq func(x) ) \geq 1 - \delta
\end{split}
\end{equation}
\normalsize}
To generate UAP, the authors used an iterative algorithm in which iteration $t>0$, a data $x$ is sampled from $\mathcal{X}$ and $\xi^u_{t-1}$ is added to the data $x$. Then, the perturbation $r$ is calculated using the following equation $(\xi^u_0 = 0)$.
\footnotesize
\begin{equation}
\begin{split}
&r_{t} = \underset{r}{\mathrm{arg\,min}} \;\left\lVert r \right\rVert_2 \\
s.t.\;& f(x+\xi^u_{t-1} + r) \neq f(x)  .
\end{split}
\end{equation}
\normalsize
To calculate $\xi^u_{t}$, $r_t$ is added to $\xi^u_{t-1}$ in each iteration, and a projection function $\mathcal{P}_{p,\epsilon}(v)$ is used which maps the input $v$ to a $L_p$ ball of radius $\epsilon$ and is centered at zero to ensure that the constraint  $\left\lVert \xi^u \right\rVert_p < \epsilon$ is satisfied. Therefore:
\footnotesize
\begin{equation}
\begin{split}
& \xi^u_{t} = \mathcal{P}_{p,\epsilon}(\xi^u_{t-1} + r_{t} ).\\
\end{split}
\end{equation}
\normalsize
The authors have demonstrated that there is no need to have much data to generate universal adversarial perturbation, and this perturbation is not unique. In another work, Brown \textit{et al.} \cite{DBLP:journals/corr/abs-1712-09665} propose universal adversarial image patch attack, which injects a patch into an image and causes the classifier to predict wrongly.



%% file: content/related.tex
There are a few studies that use adversarial machine learning approaches to evade ML-based network traffic classifiers. Verma et al. \cite{DBLP:conf/milcom/VermaCSCS18} utilize Carlini and Wanger adversarial example generating method \cite{DBLP:conf/sp/Carlini017} and a fully connected neural network as the target classifier. They use packet size and inter-arrival time features to train the target classifier for traffic characterization. Usama et al. \cite{DBLP:conf/iwcmc/UsamaQQA19} propose a new method for generating adversarial examples using Mutual Information (MI) criterion. They use MI to discover the most discriminative features to perturb them. The authors consider a support vector machine and a fully connected neural network as the target classifiers for traffic characterization. The target classifiers are trained on more than 80 statistical features for each flow, such as IAT mean, PSH flag count, forward packets per second. Previous studies \cite{DBLP:conf/milcom/VermaCSCS18,DBLP:conf/iwcmc/UsamaQQA19} do not discuss the magnitude of overhead that their methods impose on network traffic, and their methods are not universal. Imani et al. \cite{imani2019mockingbird} introduce Mockingbird to evade website fingerprinting classifiers. They focus on Deep Fingerptinitng (DF) \cite{DBLP:conf/ccs/SirinamIJW18} as the target classifier. The sequence of packets direction called trace is used as the input of DF.
Mockingbird starts with selecting a source and a target trace with different classes. It then gradually changes the source trace to become close to the target trace until the detector is deceived. Mockingbird decreases the accuracy of DF to 35.2\% with almost 56\% bandwidth overhead. Mockingbird is not universal and only works on flow time series classifiers.

The DL-based network traffic classification has been widely studied in the literature. A brief review of previous studies is provided in separate subsections and is summarized in Table \ref{tab:RW}.

\input{content/chart/RL_table}

\subsection{Packet Classification}

Lotfollahi \textit{et al.} present Deep Packet framework that embeds SDAE and CNNs to classify  ISCXVPN2016 network traffic dataset\cite{DBLP:journals/soco/LotfollahiSZS20}. 
 The Deep Packet uses the byte sequence of the transport layer header and the payload of packets to classify packets. \comment{This work is the first try to perform both application identification and traffic characterization by a DL-based classifier.} The results show that 1D-CNN classifier has better overall performance than SDAE in both application identification and traffic characterization, and the overall accuracy of 1D-CNN in application identification is 98\%, while the precision in traffic characterization is 93\%.
Wang \textit{et al.} introduce a DL-based encrypted traffic classifier called Datanet for better management of distributed smart home networks
 \cite{DBLP:journals/access/WangYCQ18}. In this study, after packets pre-processing and removing the header of the Data-link layer, the byte sequence of packets are fed into Datanet.
Datanet uses ISCXVPN2016 dataset and gains F-measure $\ge$ 96\% for both the SDAE and CNN in distinguishing 15 applications.

\subsection{Flow Content Classification}
The first attempt to apply deep learning in network traffic classification is reported by Z. Wang \textit{et al.} \cite{wang2015applications}. They recognize remarkable feature engineering ability of deep learning and try to detect protocols in a TCP flow dataset by means of SDAE. Their dataset consists of 300k pre-processed records in which exist 1000 bytes of TCP payload. They achieve more than 90\% recall in protocol detection for all protocols.
W. Wang \textit{et al.} propose an end-to-end encrypted traffic classifier to traffic characterization \cite{DBLP:conf/isi/WangZWZY17}. They use ISCXVPN2016 dataset and extract only 728 bytes of each flow as the input. 
They achieve the best result in the classification of the ISCXVPN2016 dataset compared to previous works that use the classical ML algorithms. 

After several studies in the domain of mobile application identification \cite{DBLP:journals/tifs/TaylorSCM18,DBLP:conf/eurosp/TaylorSCM16}, Aceto \textit{et al.} address mobile application identification through the Deep Learning approach for the first time\cite{DBLP:journals/tnsm/AcetoCMP19}. They investigate type of input data that is fed into the DL-classifier and consider three types of input data: ($i$) first N bytes of payload in a flow, ($ii$) first N bytes of all headers and payloads in a flow, and ($iii$) source and destination ports, number of bytes in the transport-layer header, TCP window size, inter-arrival time, and direction of packets in a flow. 
As the best result, they achieve 83\% overall accuracy for identifying 45 applications in both Android and iOS operating systems on the first type of input data. Also, they demonstrate that the overall performance of 1D-CNN and 2D-CNN are close in mobile application identification and suggest that traffic should be extracted by naturally considering data as one-dimensional. In a similar work, Rezaei \textit{et al.} propose a new DL-based classifier to identify network traffic of mobile applications \cite{8941027}. Their proposed classifier only needs the payload and statistical features of the first few packets of flows and achieves between 84\% to 98\% overall accuracy for the identification of 80 popular mobile applications.

\subsection{Flow Time Series Classification}
In \cite{DBLP:conf/icissp/Draper-GilLMG16}, a set of time-related features such as duration of flow, byte per second, and inter-arrival times between packets is utilized to  network traffic classification. They use C4.5 and K-nearest neighbors as classification techniques and achieve about 80\% accuracy in the characterization of network traffics. Moreover, they generate and distribute a labeled dataset of encrypted network traffic is called ISCXVPN2016, which has become a criterion in network traffic classification.
In \cite{DBLP:journals/cn/Caicedo-MunozEC18}, authors improve the performance of \cite{DBLP:conf/icissp/Draper-GilLMG16}, and achieve about 83\% accuracy on ISCXVPN2016 dataset using bagging and boosting classifiers. 
Lopez-Martin \textit{et al.} \cite{DBLP:journals/access/MartinCSL17} introduce a new hybrid classifier based on CNN and RNN. 
In this work, each flow comprises 20 packets. 
Regarding their results, whenever an RNN is combined with a CNN, success rate slightly improves.

The process of identifying network traffic of visited websites through privacy-enhancing technologies like Tor, which is also known as Website Fingerprinting (WF), has a background as long as encrypted traffic classification. 
Abe and Goto investigate the effectiveness of DL-based classifiers in WF for the first time \cite{abe2016fingerprinting}. In their experiment, network traffic of 100 websites that passed through Tor is monitored, and the statistical feature of each flow is extracted. They reveal that SDAE is useful in the detection websites only using direction and inter-arrival times of cells (Tor packets) with 86\% success rate. 
Rimmer \textit{et al.} indicate that DL is an effective tool in automating the process of features engineering, and 
their SDAE achieves 95.3\% success rate only using direction of Tor cells \cite{DBLP:conf/ndss/RimmerPJGJ18}. 
In another work, Sirinam \textit{et al.} design a deeper CNN classifier to outperform earlier studies, which increase the success rate to 98\% \cite{DBLP:conf/ccs/SirinamIJW18}. 

%% file: content/chart/RL_table.tex
\small
\begin{table*}[!t]
	\large
	\caption{Overview of previous studies.}
	\label{tab:RW}
	\setlength\extrarowheight{1pt}
	\resizebox{\linewidth}{!}{%
		\begin{tabular}{|c|ccccc|}
			\hline
			\rowcolor[HTML]{EFEFEF} 
			\begin{tabular}[c]{@{}c@{}}Category\end{tabular} & Paper                                                                 & Input data                  & Classifier        & Classification Target             & Year \\ \hline
			\cellcolor[HTML]{EFEFEF}                                       & P. Wang \textit{et al.} \cite{DBLP:journals/access/WangYCQ18}                 & $H^{IP}$,$H^{TL}$,$payload$             & SDAE, MLP, CNN   & Application identification        & 2018 \\
			\multirow{-2}{*}{\cellcolor[HTML]{EFEFEF}PC}                   & M. Lotfollahi \textit{et al.} \cite{DBLP:journals/soco/LotfollahiSZS20}        & $H^{TL}$,$payload$                 & SAE, CNN       & \begin{tabular}[c]{@{}c@{}}Application identification,\\ Traffic characterization \end{tabular}  & 2020 \\ \hline
			\cellcolor[HTML]{EFEFEF}                                       & Z. Wang \textit{et al.} \cite{wang2015applications}            & $H^{TL}$,$payload $                  & SAE               & Protcol detection                 & 2015 \\
			\multirow{-2}{*}{\cellcolor[HTML]{EFEFEF}FCC} 
			\cellcolor[HTML]{EFEFEF}                                       & W. Wang \textit{et al.} \cite{DBLP:conf/isi/WangZWZY17}                     & $H^{IP}$,$H^{TL}$,$payload$                 & CNN               & Traffic characterization          & 2017 \\ \hline
			\cellcolor[HTML]{EFEFEF}               & S. Rezaie \textit{et al.} \cite{8941027}                & \begin{tabular}[c]{@{}c@{}}$H^{IP}$,$H^{TL}$,$payload$,\\ $IAT$,$PS$,$Dir$\end{tabular}            & CNN, LSTM              & Application identification        & 2020 \\ 
			\multirow{-2}{*}{\cellcolor[HTML]{EFEFEF}FCC-FTSC}    & G. Aceto \textit{et al.} \cite{DBLP:journals/tnsm/AcetoCMP19}                &\begin{tabular}[c]{@{}c@{}}$H^{IP}$,$H^{TL}$,$payload$,\\ $IAT$,$PS$,$Dir$\end{tabular}                    & SAE, CNN          & Application identification        & 2019 \\ \hline
			\cellcolor[HTML]{EFEFEF}                                       & J. Caicedo-Munoz \textit{et al.} \cite{DBLP:journals/cn/Caicedo-MunozEC18}        & Time-related features       & Bagging, Boosting & Traffic characterization          & 2018 \\ 
			\cellcolor[HTML]{EFEFEF}                                       & G. Draper-Gi \textit{et al.} \cite{DBLP:conf/icissp/Draper-GilLMG16} & Time-related Features       & KNN, C4.5         & Traffic characterizaion           & 2016 \\
			\cellcolor[HTML]{EFEFEF}                                       & M. L{\'{o}}pez Mart{\'{\i}}n \textit{et al.} \cite{DBLP:journals/access/MartinCSL17}           & $PS$,$Dir$,$H^{TL}$  & LSTM, CNN          & Protcol detection                 & 2017 \\
			\cellcolor[HTML]{EFEFEF}                                       & K. Abe And S. Goto \cite{abe2016fingerprinting}                             & $Dir$,$IAT$         & SDAE              & Website fingerprinting          & 2016 \\
			\cellcolor[HTML]{EFEFEF}                                       & V. Rimmer \textit{et al.} \cite{DBLP:conf/ndss/RimmerPJGJ18}                        &  $Dir$     & SDAE, CNN, LSTM    & Website fingerprinting            & 2018 \\
			\multirow{-6}{*}{\cellcolor[HTML]{EFEFEF}FTSC}                                         & P. Sirinam \textit{et al.} \cite{DBLP:conf/ccs/SirinamIJW18}                      &  $Dir$     & CNN               & Website fingerprinting            & 2018 \\ \hline

		\end{tabular}
	}
\end{table*}
\normalsize

%% file: content/proposed_approach.tex
Adversarial Network Traffic (ANT) evaluates the robustness of DL-based network traffic classifiers, using the concept of adversarial example attack.
In adversarial example attack, an adversary adds a little perturbation into the input data and causes a classifier to make a false prediction. In the same approach to deceive DL-based network traffic classifiers, ANT injects a perturbation into network traffic.
Most of adversarial example attacks have been proposed in the context of image processing, and they are not directly applicable in the context of network traffic classification. There are some constraints to apply adversarial example attacks in the context of network traffic classification.

\begin{enumerate}
	\item 
	The content of packets must be preserved. If the content of a packet is modified, the functionality of the application that uses the packet is disrupted. Hence, we are not allowed to perturb the content of packets. The perturbation can only be injected into some specific parts of the input, such as the end of packets, or the content of a dummy packet.
	
	\item 
	In conventional machine learning tasks, like image classification, adversarial perturbation is made based on the entire data. However,  two entities make entire data in the traffic classification task, and neither of them has access to the entire data.
	
	\item
    In some network traffic types, like chat, email, and VoIP, the application that generates network traffic is not aware of the content of network traffic at the beginning of a flow, and the content of network traffic is dependent on the user. Hence,  there is no access to the entire flow when the initial packets are being processed.
	
	
	\item 
	Making a perturbation for each packet or flow has high computational overhead. 
\end{enumerate}

Based on such constraints, ANT uses Universal Adversarial Perturbation (UAP) to cause the classifier to predict wrongly. In this approach, there is no need to have access to the entire new data to make adversarial perturbation, and there is no need to compute one perturbation for each input. Also, UAP is independent of the content of target data and is made on a pre-collected set of data. To build UAP, first, a set of flows or packets of the class from which we want to make UAP is collected, and after crafting UAP, it is injected into the new incoming network traffic of that class. For example, at the first step, VoIP traffic are collected, then a UAP is made using a pre-collected set of VoIP traffic. Afterwards, when a new VoIP flow or packet is being sent, UAP is injected into it.
 Based on the category of input space, the method of generating ANT and applying UAP is different. Hence, we propose three new attacks to evaluate the robustness of network traffic classifiers which are explained afterwards.


\subsection{Adversarial Pad Attack}
We design adversarial pad (AdvPad) attack to reduce the performance of two packet classifiers PC-HP, and PC-P. AdvPad injects a UAP into the specific location of packets payload. We only consider the start and the end of the payload of packets for injecting adversarial pad and call them Start\_AdvPad and End\_AdvPad Attacks, respectively. If UAP injects into the start of the payload, the structure of an adversarially padded packet $P_i$ is as follows:
\footnotesize
\begin{equation}
\begin{split}
\textrm{PC-HP Input for $P_i$:} &<H^{TL}_{P_i}, AdvPad,Pay_{P_i}>, \\
\textrm{PC-P Input for $P_i$:} &<\!AdvPad,Pay_{P_i}>.
\end{split}
\end{equation}
\normalsize
If UAP injects into the end of the payload, the structure of an adversarially padded packet $P_i$ is as follows:
\footnotesize
\begin{equation}
\begin{split}
\textrm{PC-HP Input for $P_i$:} &<H^{TL}_{P_i},Pay_{P_i}, AdvPad>, \\
\textrm{PC-P Input for $P_i$:} &<\!Pay_{P_i},AdvPad>.
\end{split}
\end{equation}
\normalsize
\input{content/chart/advpad_alg}
 Algorithm \ref{alg:AdversarialPadding} shows the process of making UAP for the AdvPad attack. 
 This algorithm has T iterations, which in each iteration, a batch of packets are sampled from packets set $P$. All packets in $P$ has the same label $l$. 
 For $i^{th}$ packet in the batch of packets $p^{batch}$, UAP $\xi$ is injected into the start or the end of the payload, which the UAP location is specified by $Loc_{AdvPad}$ parameter. 
 A bandwidth overhead parameter $OH$ is considered to control the bandwidth overhead of  AdvPad attack.
 $OH$ determines the percentage of bandwidth overhead that we want to add to each packet. Based on this parameter, we add the first $pad\_size$ bytes of $\xi$ to a packet.
 In each iteration, the perturbation $\xi$ is updated 
 based on
 \comment{as follows: 
 \small
\begin{equation}
\begin{split}
& \Delta\xi = \epsilon \times \nabla_\xi J(\theta,packets\_byte\_sequence^{batch}, l) ,	\\
& \xi = Clip_{Domain(P)}(\xi + \Delta\xi),
\end{split}
\end{equation}
\normalsize
where $\nabla_\xi J(\theta, packets\_byte\_sequence^{batch}, l)$ is} the gradient of loss function with respect to $\xi$, and $\epsilon$ controls the magnitude of changes in $\xi$. 
The function $Clip_{Domain(P)}(x)$ maps the input $x$ to the packets domain $Domain(P)$ by clipping features that are not included. 

\subsection{Adversarial Payload Attack}
Adversarial payload (AdvPay) attack aims to deceive two flow content classifiers FCC-HP, and FCC-P, through adding UAP to the payload of a dummy packet. 
In AdvPay attack, a dummy packet is injected into a specified location $k \in [0,n-1]$ among the first $n$ packets of a flow, and UAP is injected into it. The direction of the dummy packet can be arbitrary; nonetheless, we choose the direction of the last packet before the dummy Packet $D_{p^i_{k-1}}$ as the direction of the dummy packet.
The input of  FCC-HP, and FCC-P  for flow $F_i$ are changed as follows:
\footnotesize
\begin{equation}
\begin{split}
\label{FCC-input}
\textrm{FCC-HP Input for $F_i$:}& <H^{TL}_{P^i_0}\times D_{p^i_{0}},Pay_{P^i_0}\times D_{p^i_{0}}, ...,\\
H^{TL}_{P^i_k}\times D_{p^i_{k-1}}&,AdvPay_{P^i_k}\times D_{p^i_{k-1}},\\
 ..., H^{TL}_{P^i_{n-2}}\times& D_{p^i_{n-2}},Pay_{P^i_{n-2}}\times D_{p^i_{n-2}}>, \\
\textrm{FCC-P Input for $F_i$:}& <Pay_{P^i_0}\times D_{p^i_{0}}, ..., \\
AdvPay_{P^i_k}\times D_{p^i_{k-1}},& ..., Pay_{P^i_{n-2}}\times D_{p^i_{n-2}}> .
\end{split}
\end{equation}
\normalsize

\input{content/chart/advpay_alg}

Algorithm \ref{alg:AdversarialPayload} generates UAP that is used as an adversarial payload in this attack.
In FCC-HP, the flow set $F$ consists of the transport layer header and the payload of packets, and in FCC-P, it only consists of the payload of packets.
Dummy packet index vector $IND_{AdvPay}$ is considered to determine the location of the dummy packets being injected to the flow set $F$. The $i^{th}$ element of $IND_{AdvPay}$ determines the index of the dummy packet being injected into the $i^{th}$ flow of the flow set $F$. For example, it can be the first or the second packet from source to destination or any other location in the first n packets of a flow. 
Algorithm \ref{alg:AdversarialPayload} has $T$ iterations, and in each iteration, $batch\_size$ flows from the flow set $F$ is sampled $F^{batch}$ and then the index of dummy packets of flows in $F^{batch}$ is assigned to $IND^{batch}$. For flow $i$ in $F^{batch}$,  the place of dummy packet is determined by $IND^{batch}_i$ and a dummy packet that carries UAP $\xi$ with size $AdvPay_{Size}$ bytes is injected into that place in the byte sequence of flow $f\_byte\_seq^{batch}_i$.
In each iteration, $\xi$ is updated
based on
\comment{ as follows:
\small
\begin{equation}
\begin{split}
& \Delta\xi = \epsilon \times \nabla_\xi J(\theta, flows\_byte\_seq^{batch}, l) ,		\\	
& \xi = Clip_{Domain(F)}(\xi + \Delta\xi),
\end{split}
\end{equation}
\normalsize
where $\nabla_\xi J(\theta, flow\_byte\_seq^{batch}, l) $ is} the gradient of loss function with respect to $\xi$ and $\epsilon$ tunes the size of change in $\xi$. 
The function $Clip_{Domain(F)}$ maps $\xi$ to the domain of byte sequence of flows by clipping the value of $\xi$ and controls the direction sign of dummy packet.

\subsection{Adversarial Burst Attack}
The flow time series contains the statistical features of packets in the order which they are received. A burst in a flow is a sequence of consecutive packets in one direction. The inputs FTSC-IAT and FTSC-PS for flow $F_i$ can be expressed using the concept of bursts as follows:
\footnotesize
\begin{equation}
\label{eq:FTSC_data_burst}
\begin{split}
\textrm{FTSC-IAT and FTSC-PS Inputs for $F_i$:} &\\
<(Brs^i_0,D^i_0), (Brs^i_1, D^i_1),& ..., (Brs^i_n,D^i_n)> \\
s.t. \; D^i_j \in\{-1,+1\}, \;D^i_j =  -1 \times& D^i_{j+1},\; j \in [0,n-1],
\end{split}
\end{equation}
\normalsize
where $n$ is the number of bursts in flow $F_i$ and $(Brs^i_j,D^i_j)$ is defined as follows:
\small
\begin{equation}
\label{eq:FTSC_burst}
\begin{split}
\textrm{$(Brs^i_j,D^i_j)$ for FTSC-PS:} &\\
<PS_{p^j_0}\times D_{p^j_0}, PS_{p^j_1}\times D_{p^j_1},& ..., PS_{p^j_m}\times D_{p^j_m}>, \\
\textrm{$(Brs^i_j,D^i_j)$ for FTSC-IAT:} &<IAT_{(p^{(j-1)}_{m'},p^j_0)} \times D_{p^j_0},\\
 IAT_{(p^j_0,p^j_1)} \times D_{p^j_1}, ..., &IAT_{(p^j_{m-1},p^j_m)} \times D_{p^j_m}>, \\
s.t. \; D_{p^j_k} \in\{-1,+1\}, &\;D_{p^j_k} = \; D^i_j,\; k \in [0,m],
\end{split}
\end{equation}
\normalsize
where $m$ is the number of packets in the $Brs^i_j$, $p^{(j-1)}_{m'}$ is the last packet of $Brs^i_{(j-1)}$ and if $j = 0$ then the $IAT_{(p^{(j-1)}_{m'},p^j_0)}$ will be 0. In Adversarial Burst (AdvBurst) attack, multiple dummy packets with crafted statistical features are added to the end of a selected burst of a flow. First, a burst $Brs^i_s$ from the flow $F_i$ is selected, and then, $d$ dummy packets are appended to the end of it. The direction of dummy packets and the selected burst $Brs^i_s$ is the same.
\input{content/chart/advburst_alg}
The sequence of dummy packets $<p^{dummy}_1, p^{dummy}_2, ..., p^{dummy}_d>$ and the selected burst $Brs^i_s$ build a new burst which is called adversarial burst  $Brs^i_{adv}$.
After applying the AdvBurst attack, the structure of flow $F_i$ is as follows:
\footnotesize
\begin{equation}
\label{eq:FTSC_data_burst}
\begin{split}
\textrm{$F_i$:} &<(Brs^i_0,D^i_0), (Brs^i_1, D^i_1), ..., \\
(Brs^i_{s-1},&D^i_{s-1}), (Brs^i_{adv},D^i_{s}), ..., (Brs^i_n,D^i_n)> ,
\end{split}
\end{equation}
\normalsize
where $(Brs^i_{adv},D^i_{s})$ for FTSC-IAT and FTSC-PS is defined as follows:
\footnotesize
\begin{equation}
\label{eq:FTSC_burst}
\begin{split}
\textrm{$(Brs^i_{adv},D^i_{s})$ for FTSC-PS:} &<PS_{p^s_0}\times D_{p^s_0}, ...,  PS_{p^s_m}\times\\
 D_{p^s_m}, PS_{p^{dummy}_0}&\times D^i_s, ..., PS_{p^{dummy}_d}\times D^i_s>, \\
\textrm{$(Brs^i_{adv},D^i_{s})$ for FTSC-IAT:} &<IAT_{(p^{(s-1)}_{m'},p^s_0)} \times D_{p^s_0},\\
 ..., IAT_{(p^s_{m-1},p^s_m)} \times D_{p^s_m}, &IAT_{(p^{s}_{m},p^{dummy}_0)} \times D^i_s, ...,\\ IAT_{(p^{dummy}_{d-1},p^{dummy}_d)}& \times D^i_s,> ,
\end{split}
\end{equation}
\normalsize
where $Brs^i_s$ is the selected burst, and $m$ is the number of packets in the selected burst. Algorithm \ref{alg:AdversarialBurst} generates UAP that AdvBurst attack uses as the statistical features of dummy packets. 
To specify the place of adversarial burst in each flow, this algorithm gets the selected burst indices vector $IND_{SBurst}$, which determines the indices of the selected burst in each flow.
Algorithm \ref{alg:AdversarialBurst} runs in $T$ iterations. In each iteration, $batch\_size$ number of flows are sampled $F^{batch}$, and the selected burst indices of these flows is assigned to $IND^{batch}$. For flow $i$ in $F^{batch}$, first, the flow is transformed into burst sequences $flows\_time\_series^{batch}_i$, and then the selected burst $Brs_s^i$ is replaced by the adversarial burst $Brs_{adv}^i$, which the index of selected burst $Brs_s^i$ is determined by $IND^{batch}_i$. The number of dummy packets that are added to the end of the selected burst is specified by $Num\_of\_dummy\_pkts$ parameter. The parameter $\xi$ is updated according to \comment{the following equations:
\small
\begin{equation}
\begin{split}
&	\Delta\xi = \epsilon \times \nabla_\xi J(\theta, flows\_time\_series^{batch}, l), \\		
&\xi = Clip_{Domain(F)}(\xi + \Delta\xi),
\end{split}
\end{equation}
\normalsize

where $\nabla_\xi J(\theta, flows\_time\_series^{batch}, l)$ is} the gradient of loss function with respect to $\xi$, and $\epsilon$ controls the rate of change in $\xi$. 
The function $Clip_{Domain(F)}$ makes sure that the perturbation $\xi$ is in the domain of statistical features of flows, and the sign of $\xi$ is the same as the direction of dummy packets.

%% file: content/chart/advpad_alg.tex
\begin{algorithm}[]
		\footnotesize
		\caption{ Adversarial Pad}\label{alg:AdversarialPadding}
		\hspace*{\algorithmicindent} \textbf{Input} Packet set $P$ of class $l$,  packet classifier $f_{1D-CNN}$ with \hspace*{\algorithmicindent} weights $\theta$, location of UAP $Loc_{AdvPad}$, overhead percentage $OH$, \hspace*{\algorithmicindent}\hspace{0.15cm}perturbation rate $\epsilon$, number of iterations $T$, batch size $batch\_size$. \\
		\hspace*{\algorithmicindent} \textbf{Output} UAP $\xi$
		\begin{algorithmic}[1]
			\State $\xi\gets Rand(Domain(P),Size=MaxPktSize)$
			\For{$t\gets 0, T$}
			\State $P^{batch} \gets sample\;  batch\_size\; packets\; from\; P$
			\For{$i\gets 0, batch\_size$}	
			\State $pad\_size \gets sizeof(p^{batch}_i) \times OH/100$
			\State $\xi' \gets \xi[0:pad\_size]$
			\If {$Loc_{AdvPad} == Start\;and\;func == PC-HP$}
			\State $pkt\_byte\_seq^{batch}_i \gets <H^{TL}_{p^{batch}_i},\xi',Pay_{p^{batch}_i}>$
			\ElsIf {$Loc_{AdvPad} == Start\;and\;func == PC-P$}
			\State $pkt\_byte\_seq^{batch}_i \gets <\xi',Pay_{p^{batch}_i}>$
			\ElsIf {$Loc_{AdvPad} == End\;and\;func == PC-HP$}
			\State $pkt\_byte\_seq^{batch}_i \gets <H^{TL}_{p^{batch}_i},Pay_{p^{batch}_i},\xi'>$
			\ElsIf {$Loc_{AdvPad} == End\;and\;func == PC-P$}
			\State $pkt\_byte\_seq^{batch}_i \gets <Pay_{p^{batch}_i},\xi'>$
			\EndIf
			\EndFor
			\State $\Delta\xi \gets \epsilon \times \nabla_\xi J(\theta, pkt\_byte\_seq^{batch}, l) $		
			\State $\xi \gets Clip_{Domain(P)}(\xi + \Delta\xi)$
			\EndFor
			\State \textbf{return} $\xi$
		\end{algorithmic}
\end{algorithm}

%% file: content/chart/advpay_alg.tex
\begin{algorithm}[]
		\footnotesize
		\caption{ Adversarial Payload}\label{alg:AdversarialPayload}
		\hspace*{\algorithmicindent}\textbf{Input} Flow set $F$ of class $l$,  flow content classifier $f_{1D-CNN}$ \hspace*{\algorithmicindent}with weights $\theta$, dummy packet index vector $IND_{Advpay}$, UAP size  \hspace*{\algorithmicindent}$AdvPay_{Size}$, perturbation rate $\epsilon$, number of iterations $T$, batch size \hspace*{\algorithmicindent}$batch\_size$. \\ \hspace*{\algorithmicindent}\textbf{Output} UAP $\xi$
		\begin{algorithmic}[1]
			\State $\xi\gets zeros(Size=AdvPay_{Size})$
			\For{$t\gets 0, T$}
			\State $F^{batch} \gets sample\;  batch\_size\; flows\; from\; F$
			\State $ IND^{batch} \gets dummy\; packet\; index\; of\; F^{batch}\; in\; IND_{Advpay}$
			\For{$i\gets 0, batch\_size$}	
			\State $f\_byte\_seq^{batch}_i\gets\; <>$		
			\For{$j\gets 0, n-1$}
			\If {$j < IND^{batch}_i$}
			\State$append( f\_byte\_seq^{batch}_i,< P_j^{F^{batch}_i}\times D_j^{F^{batch}_i} >)$				
			\ElsIf{ $j == IND^{batch}_i$}	
			\State $append(f\_byte\_seq^{batch}_i,< 	\hspace*{\algorithmicindent}	\hspace*{\algorithmicindent}	\hspace*{\algorithmicindent}	\hspace*{\algorithmicindent}dummy\; packet\; with\; payload\; \xi\times D_{j-1}^{F^{batch}_i}>)$		
			\ElsIf{ $j > IND^{batch}_i$}	
			\State $append(f\_byte\_seq^{batch}_i,< P_{j-1}^{F^{batch}_i}\times D_{j-1}^{F^{batch}_i}>)$			
			\EndIf
			\EndFor
			\EndFor
			\State $\Delta\xi = \epsilon \times \nabla_\xi J(\theta, f\_byte\_seq^{batch}, l) $			
			\State $\xi = Clip_{Domain(F)}(\xi + \Delta\xi)$
			\EndFor
			\State \textbf{return} $\xi$
		\end{algorithmic}
\end{algorithm}

%% file: content/chart/advburst_alg.tex
\begin{algorithm}[]
		\footnotesize
		\caption{Adversarial Burst}\label{alg:AdversarialBurst}
		\hspace*{\algorithmicindent} \textbf{Input} Flow set $F$ of class $l$,  flow content classifier $f_{1D-CNN}$ with \hspace*{\algorithmicindent}weights $\theta$, selected burst indices vector $IND_{SBurst}$,  number of dummy  \hspace*{\algorithmicindent}packets $Num\_of\_dummy\_pkts$, perturbation rate $\epsilon$, number of \\ \hspace*{\algorithmicindent}  iterations $T$,  batch size $batch\_size$. \\
		\hspace*{\algorithmicindent} \textbf{Output} UAP $\xi$
		\begin{algorithmic}[1]
			\State $\xi\gets Rand(Domain(F),Size=Num\_of\_dummy\_pkts)$
			\For{$t\gets 0, T$}
			\State $F^{batch} \gets sample\;  batch\_size\; flows\; from\; F$
			\State $ IND^{batch} \gets selected\; burst\; index\; of\; F^{batch}\; in\; IND_{SBurst}$
			\For{$i\gets 0, batch\_size$}	
			\State $flows\_time\_series^{batch}_i\gets flow\_to\_burst\_seq(F^{batch}_i)$		
			\State $Brs_s^i = flows\_time\_series^{batch}_i[IND^{batch}_i]$		
			\State $Brs_{adv}^i = append(Brs_s^i,\xi)$	
			\State $flows\_time\_series^{batch}_i[IND^{batch}_i] = Brs_{adv}^i$		
			\EndFor
			\State $\Delta\xi = \epsilon \times \nabla_\xi J(\theta, flows\_time\_series^{batch}, l) $			
			\State $\xi = Clip_{Domain(F)}(\xi + \Delta\xi)$
			\EndFor
			\State \textbf{return} $\xi$
		\end{algorithmic}
\end{algorithm}

%% file: content/evaluation.tex
Three attacks have been proposed in order to evaluate the robustness of the DL-based network traffic classifiers with respect to the three categories of input space.
In this section, the robustness of six different classifiers is examined against ANT. 
Moreover, the robustness of classifiers that get the transport layer header as the input, PC-HP and FCC-HP, are evaluated against port attack. In port attack, source and destination port numbers of flows are randomly assigned from a specific range of port numbers. 
Typical performance metrics for evaluating a classifier are precision, recall, and F-score. 
The recall shows what fraction of given class data is classified correctly.
Since proposed attacks cause the classifiers to misclassify data of a given class, 
we focus on recall in the evaluation section.
In the following experiments, we assume an adversary  has white-box access to DL-based classifiers. In the white-box setting, an adversary (\textit{e.g.,} malicious cloud provider) knows the input space category, the architecture, the weights, and the outputs of the classifier. Papernot \textit{et al.} \cite{DBLP:journals/corr/PapernotMG16} have shown that adversarial examples are transferable among different classifiers with different training data set. In the black-box setting, an adversary can use proposed attacks and run them on a substitute classifier, which is close to the target classifier, and he/she uses the generated perturbations to evade a target classifier.
We evaluate the transferability of ANT in section \ref{sec:blackbox}.

\subsection{Dataset}
DL-based network traffic classifiers need to be trained using a labeled set of network traffic in the training phase. 
To the best of our knowledge, a few studies have a reliable labeling process, and some of them have not enough data to train a deep neural network. 
Draper \textit{et al.} \cite{DBLP:conf/icissp/Draper-GilLMG16} have presented ISCXVPN2016 dataset that consists of six types of network traffic, and some of flows in the dataset have been sent through a VPN. They used different applications, such as Skype, Facebook, Hangouts, Youtube, and Bittorent, to generate traffic with different types, including chat, email, file transfer, streaming, torrent, and VoIP. This dataset has been labeled based on application and type, which have generated given network traffic. 
This dataset is highly imbalanced and 
it decreases the performance of DL-based classifiers. We use undersampling for
classes having more data and oversampling for classes having fewer data to overcome this challenge \cite{DBLP:journals/csur/BrancoTR16}.
	 The undersampling method eliminates a certain amount of examples from the training dataset that belong to the majority class. The oversampling method replicates examples from the training dataset that belong to the minority class until the amount of data for that class reaches a specific number. The dataset is split into training (60\%), validation (20\%), and test sets (20\%). The dataset requires a preprocessing phase to clean some background network traffic, such as DNS and NETBIOS. Since there are three categories of input spaces, data padding and normalization of each category is different. Hence, we discuss dataset properties of each category in a separate subsection.
 \comment{
 	
\subsection{Performance Metrics}

The precision for a given label denotes what fraction of data is correctly classified (label confidence). F-score is a combination of precision and recall and indicates how fine are both of these  metrics. These three metrics are used for data of one class. The equations are as follows:
\small
\begin{equation*}
\begin{split}
& Precision = \frac{TP}{TP + FP}, \quad  Recall = \frac{TP}{TP + FN}, \quad \\&F\!\!-\!\!score = \frac{2}{Precision^{-1}+ Recall^{-1}}.
\end{split}
\end{equation*}
\normalsize
 The overall accuracy is used as a classifier performance metric on all classes data, and the overall accuracy of a classifier with n classes is defined as follows:
 \small
\begin{equation*}
\begin{split}
&Overall  Accuracy = \frac{TP_1+ TP_2+ ... + TP_n }{Number\; of\; all\; classes\; data},
\end{split}
\end{equation*}
\normalsize
where $TP_i$ is the number of true positive data of $i^{th}$ class. 
Since proposed attacks cause the classifiers to misclassify data of a given class, the number of true-positive data of that class decreases, and the number of false-positive data of that class is constant. 
Consequently, since the denominators of recall and overall accuracy are fixed, we focus on these metrics in the evaluation section.}

\subsection{Packet Classification}
\input{content/eval_pkt_classification.tex}

\subsection{Flow Content Classification}
\input{content/eval_flow_classification.tex}

\subsection{Flow Time Series Classification}
\input{content/eval-ts-classification.tex}

\subsection{Transferability of ANT}
\label{sec:blackbox}
\input{content/blackbox}

%% file: content/eval_pkt_classification.tex
In this section, the robustness of packet classifiers against port and adversarial pad (AdvPad) attacks is evaluated. We propose a random pad attack (RandPad) as a baseline to assess the effectiveness of the AdvPad attack. RandPad injects a random pad into the payload of packets. AdvPad must have a better performance than RandPad . In the AdvPad attack, for a given class, Algorithm \ref{alg:AdversarialPadding} generates a UAP on the packets of that class in the validation set with the desired percentage of bandwidth overhead, and then the UAP is injected into packets of that class in the test set, and the performance of packet classifiers is evaluated against these packets.

\subsubsection{Packet Classification Dataset}
Packet classification dataset consists of the byte sequence of packets that have at least one byte of payload. Since the decimal domain of each byte of a packet is in the [0,255] interval, we divide the decimal value of each byte by 255 to normalize the byte sequence of packets. Hence, each packet becomes a sequence of numbers between zero and one. Also, we need to add zero pad to the end of packets until their length reaches to $MaxPktSize$.  
Table \ref{tab:PacketClassificationDataset} shows the number of samples in each class. 

\input{content/chart/pc_dataset}

\subsubsection{Packet classifiers Setup}
A 1D-CNN is used to classify the byte sequences of packets. Given our computational resources, we trained 50 1D-CNNs with different architectures and hyper-parameters, and finally, the classifier with higher accuracy on the validation set has been selected. 
The overall accuracy of PC-HP and PC-P is 88.09\% and 66.71\%, respectively. The precision, recall, and F-score of these classifiers are reported in Table \ref{tab:PC_perf_metric}. 
PC-HP has higher performance than PC-P because of the information in the transport layer header of packets. 

\input{content/chart/pc_perf}

\subsubsection{Packet Classifiers Robustness Evaluation}

\input{content/chart/pc_eval}

AdvPad, RandPad, and port attack are applied to packet classifiers to evaluate their robustness. AdvPad has been conducted for 1000 iterations with $batch\_size = 128$ and $\epsilon=0.01$. RandPad has been run 50 times, and the average results are reported in this section. 
Figure \ref{fig:pc_evala} shows the recall of PC-HP under proposed attacks for the various magnitude of BandWidth (BW) overhead for all classes. 

Port attack randomly changes the transport layer port numbers of flows, and it does not impose any BW overhead on network traffic.
The recall of chat, email, and streaming classes decreases considerably under port attack that shows PC-HP is highly sensitive to the transport layer port numbers for classifying these classes. However, the port attack has little effect on the recall of other classes.

End\_AdvPad considerably decreases the recall of PC-HP in most classes with little bandwidth overhead, and recall is more decreased by increasing the size of the adversarial pad. If it combines with port attack (End\_AdvPad+Port), the performance of attack is raised in most classes. 
End\_RandPad injects a random pad at the end of packets payload,  and End\_RandPad+Port combines End\_RandPad with port attack.
End\_AdvPad and End\_AdvPad+Port have better performance than End\_RandPad and End\_RandPad+Port attacks, respectively. This observation shows the effectiveness of the AdvPad attack. 

Start\_AdvPad reduces the recall of all classes to under 25\% with just 10\% bandwidth overhead. The recall of PC-HP under Start\_AdvPad gets close to 0\% by increasing the size of adversarial pad. Adding Port attack to Start\_AdvPad (Start\_AdvPad+Port) does not improve the performance of Start\_AdvPad in most classes.
In the Start\_RandPad attack, a random pad is injected into the start of packets payload, and Start\_RandPad+port  mixes Start\_RandPad with port attack. There is a significant gap between the recall of PC-HP under Start\_AdvPad and Start\_RandPad, and also, there is a large gap between the recall of PC-HP under Start\_AdvPad+port and Start\_RandPad+port. These gaps show the remarkable effectiveness of the adversarial pad at the start of packets for all classes.
The recall of PC-HP under Start\_AdvPad and Start\_AdvPad+Port is reduced more than the End\_AdvPad and the End\_AdvPad+Port for all classes, respectively. These observations show PC-HP is more sensitive to the information at the start of packets payload than the information at the end of it. 
The recall of PC-HP under the End\_RandPad and Start\_RandPad are very analogous to each other in some classes.
The performance of random pad attacks is lower than the End\_AdvPad and the End\_AdvPad+Port attacks in most classes. These results show the effectiveness of adversarial pad attacks.

Figure \ref{fig:pc_evalb} indicates the recall of PC-P under proposed attacks for different magnitudes of BW overhead for all classes. 
The performance of the End\_AdvPad is variable in various classes. Although the decline of recall of PC-P under the End\_AdvPad is low in file transfer and chat classes, the End\_AdvPad highly decreases the recall of PC-P in other classes.
The recall of PC-P under the End\_RandPad is a little higher than the recall of PC-P under the End\_AdvPad in most classes, which indicates low effectiveness of the End\_AdvPad on PC-P. 
Start\_AdvPad significantly decreases the recall of PC-P and indicates the best performance among all attacks on PC-P. It decreases the recall of all classes to less than 3\% with just 30\% BW overhead. Start\_AdvPad has better performance than Start\_RandPad in all classes.
In some classes such as VoIP, email, chat, and file transfer, the recall of PC-P under the Start\_RandPad is close or lower than recall of PC-P under the End\_AdvPad. This observation demonstrates PC-P is very sensitive to the information at the start of packets.

%% file: content/chart/pc_dataset.tex
\begin{table}[]
	\caption{Packet classification dataset.}
	\label{tab:PacketClassificationDataset}
	\setlength\extrarowheight{2pt}
	\resizebox{\linewidth}{!}{%
		\begin{tabular}{|
				>{\columncolor[HTML]{EFEFEF}}c |c|c|c|c|}
			\hline
			\cellcolor[HTML]{EFEFEF} & \multicolumn{2}{c|}{\cellcolor[HTML]{EFEFEF}Imbalanced Dataset} & \multicolumn{2}{c|}{\cellcolor[HTML]{EFEFEF}Balanced Dataset} \\ \cline{2-5} 
			\multirow{-2}{*}{\cellcolor[HTML]{EFEFEF}Type} & \cellcolor[HTML]{EFEFEF}{\small Total Number} & \cellcolor[HTML]{EFEFEF}{\small Training Set} & \cellcolor[HTML]{EFEFEF}{\small Total Number} & \cellcolor[HTML]{EFEFEF}{\small Training Set} \\ \hline
			Chat & 68,864 & 41,646 & 68,864 & 41,646 \\ \hline
			Email & 30,807 & 18,558 & 51,849 & 39,600 \\ \hline
			FileTransfer & 313,916 & 188,349 & 100,796 & 60,505 \\ \hline
			Streaming & 124,487 & 74,692 & 100,158 & 60,023 \\ \hline
			Torrent & 162,479 & 97,487 & 101,164 & 60,625 \\ \hline
			VoIP & 596,476 & 357,885 & 100,432 & 60,018 \\ \hline
		\end{tabular}
	}
\end{table}

%% file: content/chart/pc_perf.tex

\begin{table}[]
	\centering
	\setlength\extrarowheight{1pt}
	\caption{Performance metrics of the packet classifiers PC-HP, and PC-P.}
	\label{tab:PC_perf_metric}
	\begin{tabular}{|
			>{\columncolor[HTML]{EFEFEF}}c |cc|cc|cc|}
		\hline
		\cellcolor[HTML]{EFEFEF} & \multicolumn{2}{c|}{\cellcolor[HTML]{EFEFEF}Precision(\%)} & \multicolumn{2}{c|}{\cellcolor[HTML]{EFEFEF}Recall(\%)} & \multicolumn{2}{c|}{\cellcolor[HTML]{EFEFEF}F-score(\%)} \\ \cline{2-7} 
		\multirow{-2}{*}{\cellcolor[HTML]{EFEFEF}Type} & \cellcolor[HTML]{EFEFEF}PC-HP & \cellcolor[HTML]{EFEFEF}PC-P & \cellcolor[HTML]{EFEFEF}PC-HP & \cellcolor[HTML]{EFEFEF}PC-P & \cellcolor[HTML]{EFEFEF}PC-HP & \cellcolor[HTML]{EFEFEF}PC-P \\ \hline
		Chat & 85.35 & 74.11 & 76.45 & 47.31 & 80.66 & 57.75 \\ \hline
		Email & 65.77 & 59.97 & 75.35 & 69.84 & 70.23 & 64.53 \\ \hline
		FileTransfer & 98.41 & 67.51 & 90.23 & 77.19 & 94.14 & 72.03 \\ \hline
		Streaming & 79.96 & 48.08 & 91.50 & 41.88 & 85.34 & 44.77 \\ \hline
		Torrent & 99.3 & 69.66 & 99.97 & 77.33 & 99.63 & 73.29 \\ \hline
		VoIP & 87.24 & 76.33 & 84.47 & 85.15 & 85.83 & 80.50 \\ \hline
	\end{tabular}
\end{table}

%% file: content/chart/pc_eval.tex
\newcommand{\chatr}{pktcls_chat_recall.csv}
\newcommand{\emailr}{pktcls_email_recall.csv}
\newcommand{\filer}{pktcls_filetransfer_recall.csv}
\newcommand{\streamingr}{pktcls_streaming_recall.csv}
\newcommand{\torrentr}{pktcls_torrent_recall.csv}
\newcommand{\voipr}{pktcls_voip_recall.csv}
\newcommand{\pcend}{csv/pc_end/}
\newcommand{\pcstart}{csv/pc_start/}
\newcommand{\acccsv}{pktcls_Acc.csv}

\newcommand{\colorone}{blue}
\newcommand{\colortwo}{red}
\newcommand{\colorthree}{olive}
\newcommand{\colorfour}{teal}
\newcommand{\colorfive}{green}
\newcommand{\colorsix}{brown}

\newcommand{\markone}{triangle*}
\newcommand{\marktwo}{*}
\newcommand{\markthree}{square*}
\newcommand{\markfour}{+}
\newcommand{\markfive}{*}
\newcommand{\marksix}{x}


\newcommand{\pcpend}{csv/PC-P-End/}
\newcommand{\pcpstart}{csv/PC-P-Start/}

\begin{figure*}
	\centering
	\begin{subfigure}{\linewidth}
		\centering
		\begin{tikzpicture}
		\pgfplotsset{
			scaled y ticks = false,
			width=4cm,
			height=4cm,
			axis on top,
			xmin=-3,xmax=53,
			ymin=-0,ymax=102,
			minor y tick num=1,
			minor x tick num=1,
			label style = {font = {\fontsize{8 pt}{12 pt}\selectfont}},
			title style = {font = {\fontsize{9pt}{12 pt}\selectfont}},
			ytick={0,20,40,60,80,100},
			xtick={0,10,20,30,40,50},
			x tick label style={rotate=45},
			tick label style = {font = {\fontsize{4 pt}{12 pt}\selectfont}},
			ylabel shift={-0.8em},
			ylabel style={align=center},
			xlabel shift={-0.4em},
			grid=both, 
			grid style={dashed,gray!30}
		}
		\begin{groupplot}[ 
		group style={
			group size=7 by 1,
			vertical sep=5pt,
			horizontal sep=17pt
		},
		]
		\nextgroupplot[ xlabel={},ylabel={Recall(\%)},title={Chat}]
		\addplot[\colorone,mark=\markone,mark size=1.0]    table[x=overhead,y=rand,col sep=comma]{\pcend\chatr};
		\addplot[\colortwo,mark=\marktwo,mark size=1.0]    table[x=overhead,y=advpad,col sep=comma]{\pcend\chatr};  
		\addplot[\colorone,densely dashed ,mark=\markone,mark size=1.0]    table[x=overhead,y=rand_port,col sep=comma]{\pcend\chatr};
		\addplot[\colortwo,densely dashed ,mark=\marktwo,mark size=1.0]    table[x=overhead,y=advpad_port,col sep=comma]{\pcend\chatr};
		\addplot[\colorthree,mark=\markthree,mark size=1.0]    table[x=overhead,y=rand,col sep=comma]{\pcstart\chatr};
		\addplot[\colorfour,mark=\markfour,mark size=1.0]    table[x=overhead,y=advpad,col sep=comma]{\pcstart\chatr};  
		\addplot[\colorthree,densely dashed ,mark=\markthree,mark size=1.0]    table[x=overhead,y=rand_port,col sep=comma]{\pcstart\chatr};
		\addplot[\colorfour,densely dashed,mark=\markfour,mark size=1.0 ]    table[x=overhead,y=advpad_port,col sep=comma]{\pcstart\chatr};
		\addplot[only marks,\colorfive,mark=\markfive,mark size=1.5]    table[x=overhead,y=natural,col sep=comma]{\pcend\chatr};  
		\addplot[only marks,\colorsix,mark=\marksix]    table[x=overhead,y=port,col sep=comma]{\pcend\chatr};   
		\coordinate (c1) at (rel axis cs:0,1);
		
		\nextgroupplot[ xlabel={},ylabel={},title={Email},]
		\addplot[\colorone,mark=\markone,mark size=1.0]    table[x=overhead,y=rand,col sep=comma]{\pcend\emailr};
		\addplot[\colortwo,mark=\marktwo,mark size=1.0]    table[x=overhead,y=advpad,col sep=comma]{\pcend\emailr};  
		\addplot[\colorone,densely dashed ,mark=\markone,mark size=1.0]    table[x=overhead,y=rand_port,col sep=comma]{\pcend\emailr};
		\addplot[\colortwo,densely dashed ,mark=\marktwo,mark size=1.0]    table[x=overhead,y=advpad_port,col sep=comma]{\pcend\emailr};
		\addplot[\colorthree,mark=\markthree,mark size=1.0]    table[x=overhead,y=rand,col sep=comma]{\pcstart\emailr};
		\addplot[\colorfour,mark=\markfour,mark size=1.0]    table[x=overhead,y=advpad,col sep=comma]{\pcstart\emailr};  
		\addplot[\colorthree,densely dashed ,mark=\markthree,mark size=1.0]    table[x=overhead,y=rand_port,col sep=comma]{\pcstart\emailr};
		\addplot[\colorfour,densely dashed,mark=\markfour,mark size=1.0 ]    table[x=overhead,y=advpad_port,col sep=comma]{\pcstart\emailr};
		\addplot[only marks,\colorfive,mark=\markfive,mark size=1.5]    table[x=overhead,y=natural,col sep=comma]{\pcend\emailr};  
		\addplot[only marks,\colorsix,mark=\marksix]    table[x=overhead,y=port,col sep=comma]{\pcend\emailr};

		\nextgroupplot[
		ylabel={},title={FileTransfer},
		legend style={nodes={scale=0.8, transform shape},at={($(-,-1cm)+(-1cm,-1cm)$)},legend columns=5,fill=none,draw=black,anchor=center,align=center},
		legend to name=fred
		]
		\addplot[\colortwo,mark=\marktwo,mark size=1.0]    table[x=overhead,y=advpad,col sep=comma]{\pcend\filer};  
		\addplot[\colorone,mark=\markone,mark size=1.0]    table[x=overhead,y=rand,col sep=comma]{\pcend\filer};
		\addplot[\colortwo,densely dashed ,mark=\marktwo,mark size=1.0]    table[x=overhead,y=advpad_port,col sep=comma]{\pcend\filer};
		\addplot[\colorone,densely dashed ,mark=\markone,mark size=1.0]    table[x=overhead,y=rand_port,col sep=comma]{\pcend\filer};
		\addplot[\colorfour,mark=\markfour,mark size=1.0]    table[x=overhead,y=advpad,col sep=comma]{\pcstart\filer};  
		\addplot[\colorthree,mark=\markthree,mark size=1.0]    table[x=overhead,y=rand,col sep=comma]{\pcstart\filer};
		\addplot[\colorfour,densely dashed,mark=\markfour,mark size=1.0 ]    table[x=overhead,y=advpad_port,col sep=comma]{\pcstart\filer};
		\addplot[\colorthree,densely dashed ,mark=\markthree,mark size=1.0]    table[x=overhead,y=rand_port,col sep=comma]{\pcstart\filer};
		\addplot[only marks,\colorsix,mark=\marksix]    table[x=overhead,y=port,col sep=comma]{\pcend\filer};  
		\addplot[only marks,\colorfive,mark=\markfive,mark size=1.5]    table[x=overhead,y=natural,col sep=comma]{\pcend\filer};

		\coordinate (c2) at (rel axis cs:1,1);
		\nextgroupplot[ xlabel={},ylabel={},title={Streaming},]
		\addplot[\colorone,mark=\markone,mark size=1.0]    table[x=overhead,y=rand,col sep=comma]{\pcend\streamingr};
		\addplot[\colortwo,mark=\marktwo,mark size=1.0]    table[x=overhead,y=advpad,col sep=comma]{\pcend\streamingr};  
		\addplot[\colorone,densely dashed ,mark=\markone,mark size=1.0]    table[x=overhead,y=rand_port,col sep=comma]{\pcend\streamingr};
		\addplot[\colortwo,densely dashed ,mark=\marktwo,mark size=1.0]    table[x=overhead,y=advpad_port,col sep=comma]{\pcend\streamingr};
		\addplot[\colorthree,mark=\markthree,mark size=1.0]    table[x=overhead,y=rand,col sep=comma]{\pcstart\streamingr};
		\addplot[\colorfour,mark=\markfour,mark size=1.0]    table[x=overhead,y=advpad,col sep=comma]{\pcstart\streamingr};  
		\addplot[\colorthree,densely dashed ,mark=\markthree,mark size=1.0]    table[x=overhead,y=rand_port,col sep=comma]{\pcstart\streamingr};
		\addplot[\colorfour,densely dashed,mark=\markfour,mark size=1.0 ]    table[x=overhead,y=advpad_port,col sep=comma]{\pcstart\streamingr};
		\addplot[only marks,\colorfive,mark=\markfive,mark size=1.5]    table[x=overhead,y=natural,col sep=comma]{\pcend\streamingr};  
		\addplot[only marks,\colorsix,mark=\marksix]    table[x=overhead,y=port,col sep=comma]{\pcend\streamingr};   
		
		\nextgroupplot[ xlabel={},ylabel={},title={Torrent},]
		\addplot[\colorone,mark=\markone,mark size=1.0]    table[x=overhead,y=rand,col sep=comma]{\pcend\torrentr};
		\addplot[\colortwo,mark=\marktwo,mark size=1.0]    table[x=overhead,y=advpad,col sep=comma]{\pcend\torrentr};  
		\addplot[\colorone,densely dashed ,mark=\markone,mark size=1.0]    table[x=overhead,y=rand_port,col sep=comma]{\pcend\torrentr};
		\addplot[\colortwo,densely dashed ,mark=\marktwo,mark size=1.0]    table[x=overhead,y=advpad_port,col sep=comma]{\pcend\torrentr};
		\addplot[\colorthree,mark=\markthree,mark size=1.0]    table[x=overhead,y=rand,col sep=comma]{\pcstart\torrentr};
		\addplot[\colorfour,mark=\markfour,mark size=1.0]    table[x=overhead,y=advpad,col sep=comma]{\pcstart\torrentr};  
		\addplot[\colorthree,densely dashed ,mark=\markthree,mark size=1.0]    table[x=overhead,y=rand_port,col sep=comma]{\pcstart\torrentr};
		\addplot[\colorfour,densely dashed,mark=\markfour,mark size=1.0 ]    table[x=overhead,y=advpad_port,col sep=comma]{\pcstart\torrentr};
		\addplot[only marks,\colorfive,mark=\markfive,mark size=1.5]    table[x=overhead,y=natural,col sep=comma]{\pcend\torrentr};  
		\addplot[only marks,\colorsix,mark=\marksix]    table[x=overhead,y=port,col sep=comma]{\pcend\torrentr};   
		
		\nextgroupplot[ xlabel={},ylabel={},title={VoIP},]
		\addplot[\colorone,mark=\markone,mark size=1.0]    table[x=overhead,y=rand,col sep=comma]{\pcend\voipr};
		\addplot[\colortwo,mark=\marktwo,mark size=1.0]    table[x=overhead,y=advpad,col sep=comma]{\pcend\voipr};  
		\addplot[\colorone,densely dashed ,mark=\markone,mark size=1.0]    table[x=overhead,y=rand_port,col sep=comma]{\pcend\voipr};
		\addplot[\colortwo,densely dashed ,mark=\marktwo,mark size=1.0]    table[x=overhead,y=advpad_port,col sep=comma]{\pcend\voipr};
		\addplot[\colorthree,mark=\markthree,mark size=1.0]    table[x=overhead,y=rand,col sep=comma]{\pcstart\voipr};
		\addplot[\colorfour,mark=\markfour,mark size=1.0]    table[x=overhead,y=advpad,col sep=comma]{\pcstart\voipr};  
		\addplot[\colorthree,densely dashed ,mark=\markthree,mark size=1.0]    table[x=overhead,y=rand_port,col sep=comma]{\pcstart\voipr};
		\addplot[\colorfour,densely dashed,mark=\markfour,mark size=1.0 ]    table[x=overhead,y=advpad_port,col sep=comma]{\pcstart\voipr};
		\addplot[only marks,\colorfive,mark=\markfive,mark size=1.5]    table[x=overhead,y=natural,col sep=comma]{\pcend\voipr};  
		\addplot[only marks,\colorsix,mark=\marksix]    table[x=overhead,y=port,col sep=comma]{\pcend\voipr};   
		
		
		
		\end{groupplot}
		\coordinate (c3) at ($(c1)!.5!(c2)$);
		\node[below] at (c2 |- current bounding box.south)
		{\pgfplotslegendfromname{fred}};
		\end{tikzpicture}%
				\vspace{-2.3\baselineskip}
		\label{fig:sub1}
		\caption{ }
					\label{fig:pc_evalb}
	\end{subfigure}
	\begin{subfigure}{\linewidth}
		\centering
		\begin{tikzpicture}
		\pgfplotsset{
			scaled y ticks = false,
			width=4cm,
			height=4cm,
			axis on top,
			xmin=-3,xmax=53,
			ymin=-0,ymax=102,
			label style = {font = {\fontsize{8 pt}{12 pt}\selectfont}},
			title style = {font = {\fontsize{9pt}{12 pt}\selectfont}},
			minor y tick num=1,
			minor x tick num=1,
			ytick={0,20,40,60,80,100},
			xtick={0,10,20,30,40,50},
			x tick label style={rotate=45},
			tick label style = {font = {\fontsize{4 pt}{12 pt}\selectfont}},
			ylabel shift={-0.8em},
			ylabel style={align=center},
			xlabel shift={-0.4em},
			grid=both, 
			grid style={dashed,gray!30}
		}

		\begin{groupplot}[ 
		group style={
			group size=6 by 1,
			vertical sep=45pt,
			horizontal sep=17pt
		},
		]
		\nextgroupplot[ xlabel={},ylabel={Recall(\%)}]
		\addplot[\colorone,mark=\markone,mark size=1.0]    table[x=overhead,y=rand,col sep=comma]{\pcpend\chatr};
		\addplot[\colortwo,mark=\marktwo,mark size=1.0]    table[x=overhead,y=advpad,col sep=comma]{\pcpend\chatr};  
		\addplot[\colorthree,mark=\markthree,mark size=1.0]    table[x=overhead,y=rand,col sep=comma]{\pcpstart\chatr};
		\addplot[\colorfour,mark=\markfour,mark size=1.0]    table[x=overhead,y=advpad,col sep=comma]{\pcpstart\chatr};  
		\addplot[only marks,\colorfive,mark=\markfive,mark size=1.5]    table[x=overhead,y=natural,col sep=comma]{\pcpend\chatr};  
		\coordinate (c1) at (rel axis cs:0,1);
		
		\nextgroupplot[ xlabel={},ylabel={}]
		\addplot[\colorone,mark=\markone,mark size=1.0]    table[x=overhead,y=rand,col sep=comma]{\pcpend\emailr};
		\addplot[\colortwo,mark=\marktwo,mark size=1.0]    table[x=overhead,y=advpad,col sep=comma]{\pcpend\emailr};  
		\addplot[\colorthree,mark=\markthree,mark size=1.0]    table[x=overhead,y=rand,col sep=comma]{\pcpstart\emailr};
		\addplot[\colorfour,mark=\markfour,mark size=1.0]    table[x=overhead,y=advpad,col sep=comma]{\pcpstart\emailr};  
		\addplot[only marks,\colorfive,mark=\markfive,mark size=1.5]    table[x=overhead,y=natural,col sep=comma]{\pcpend\emailr};

		\nextgroupplot[
		xlabel={BandWidth Overhead (\%)},x label style={at={(axis description cs:1.2,-0.15)},anchor=north},ylabel={},]

		]
		\addplot[\colortwo,mark=\marktwo,mark size=1.0]    table[x=overhead,y=advpad,col sep=comma]{\pcpend\filer};  
		\addplot[\colorone,mark=\markone,mark size=1.0]    table[x=overhead,y=rand,col sep=comma]{\pcpend\filer};
		\addplot[\colorfour,mark=\markfour,mark size=1.0]    table[x=overhead,y=advpad,col sep=comma]{\pcpstart\filer};  
		\addplot[\colorthree,mark=\markthree,mark size=1.0]    table[x=overhead,y=rand,col sep=comma]{\pcpstart\filer};
		
		\addplot[only marks,\colorfive,mark=\markfive,mark size=1.5]    table[x=overhead,y=natural,col sep=comma]{\pcpend\filer};       
		\coordinate (c2) at (rel axis cs:1,1);
		\nextgroupplot[ xlabel={},ylabel={}]
		\addplot[\colorone,mark=\markone,mark size=1.0]    table[x=overhead,y=rand,col sep=comma]{\pcpend\streamingr};
		\addplot[\colortwo,mark=\marktwo,mark size=1.0]    table[x=overhead,y=advpad,col sep=comma]{\pcpend\streamingr};  
		\addplot[\colorthree,mark=\markthree,mark size=1.0]    table[x=overhead,y=rand,col sep=comma]{\pcpstart\streamingr};
		\addplot[\colorfour,mark=\markfour,mark size=1.0]    table[x=overhead,y=advpad,col sep=comma]{\pcpstart\streamingr};  
		\addplot[only marks,\colorfive,mark=\markfive,mark size=1.5]    table[x=overhead,y=natural,col sep=comma]{\pcpend\streamingr};  
		
		\nextgroupplot[ xlabel={},ylabel={}]
		\addplot[\colorone,mark=\markone,mark size=1.0]    table[x=overhead,y=rand,col sep=comma]{\pcpend\torrentr};
		\addplot[\colortwo,mark=\marktwo,mark size=1.0]    table[x=overhead,y=advpad,col sep=comma]{\pcpend\torrentr};  
		\addplot[\colorthree,mark=\markthree,mark size=1.0]    table[x=overhead,y=rand,col sep=comma]{\pcpstart\torrentr};
		\addplot[\colorfour,mark=\markfour,mark size=1.0]    table[x=overhead,y=advpad,col sep=comma]{\pcpstart\torrentr};  
		\addplot[\colorthree,densely dashed ,mark=\markthree,mark size=1.0]    table[x=overhead,y=rand_port,col sep=comma]{\pcpstart\torrentr};
		\addplot[\colorfour,densely dashed,mark=\markfour,mark size=1.0 ]    table[x=overhead,y=advpad_port,col sep=comma]{\pcpstart\torrentr};
		\addplot[only marks,\colorfive,mark=\markfive,mark size=1.5]    table[x=overhead,y=natural,col sep=comma]{\pcpend\torrentr};  
		\addplot[only marks,\colorsix,mark=\marksix]    table[x=overhead,y=port,col sep=comma]{\pcpend\torrentr};   
		
		\nextgroupplot[ xlabel={},ylabel={}]
		\addplot[\colorone,mark=\markone,mark size=1.0]    table[x=overhead,y=rand,col sep=comma]{\pcpend\voipr};
		\addplot[\colortwo,mark=\marktwo,mark size=1.0]    table[x=overhead,y=advpad,col sep=comma]{\pcpend\voipr};  
		\addplot[\colorthree,mark=\markthree,mark size=1.0]    table[x=overhead,y=rand,col sep=comma]{\pcpstart\voipr};
		\addplot[\colorfour,mark=\markfour,mark size=1.0]    table[x=overhead,y=advpad,col sep=comma]{\pcpstart\voipr};  
		\addplot[only marks,\colorfive,mark=\markfive,mark size=1.5]    table[x=overhead,y=natural,col sep=comma]{\pcpend\voipr};  
		
		
		
		\end{groupplot}
		\coordinate (c3) at ($(c1)!.5!(c2)$);
		\node[below] at (c2 |- current bounding box.south)
		{\pgfplotslegendfromname{fred}};
		\end{tikzpicture}%
		\vspace{-2.3\baselineskip}
		\label{fig:sub2}
		\caption{ }
			\label{fig:pc_evalb}
	\end{subfigure}
\begin{subfigure}{\linewidth}
	\centering
	\begin{tikzpicture} 
	\begin{axis}[%
	hide axis,
	xmin=10,
	xmax=50,
	ymin=0,
	ymax=0.4,
	legend style={nodes={scale=0.8, transform shape},at={($(-1cm,-1cm)+(-1cm,-1cm)$)},legend columns=5,fill=none,draw=black,anchor=center,align=center}
	]
	\addlegendimage{\colortwo,mark=\marktwo,mark size=1.0}
	\addlegendentry{End\_AdvPad};  
	\addlegendimage{\colorone,mark=\markone,mark size=1.0}
	\addlegendentry{End\_RandPad};     
	\addlegendimage{ \colortwo,densely dashed ,mark=\marktwo,mark size=1.0}
	\addlegendentry{End\_AdvPad+Port};
	\addlegendimage{\colorone,densely dashed ,mark=\markone,mark size=1.0}
	\addlegendentry{End\_RandPad+Port};  
	\addlegendimage{\colorfour,mark=\markfour,mark size=1.0}
	\addlegendentry{Start\_AdvPad};  	
	\addlegendimage{\colorthree,mark=\markthree,mark size=1.0}
	\addlegendentry{Start\_RandPad};

	\addlegendimage{ \colorfour,densely dashed,mark=\markfour,mark size=1.0}
	\addlegendentry{Start\_AdvPad+Port};
	\addlegendimage{\colorthree,densely dashed ,mark=\markthree,mark size=1.0}
	\addlegendentry{Start\_RandPad+Port};  
	
	\addlegendimage{only marks,\colorsix,mark=\marksix}
	\addlegendentry{Port};      
	\addlegendimage{ only marks,\colorfive,mark=\markfive}
	\addlegendentry{No Attack};    
	
	\end{axis}
	\end{tikzpicture}
\end{subfigure}

	\caption{The recall of PC-HP (a) and PC-P (b) under different attacks over various sizes of BandWidth Overhead (BWO) for all classes. The legends show various kinds of attacks that have been applied.}
	\label{fig:pc_evala}
\end{figure*}
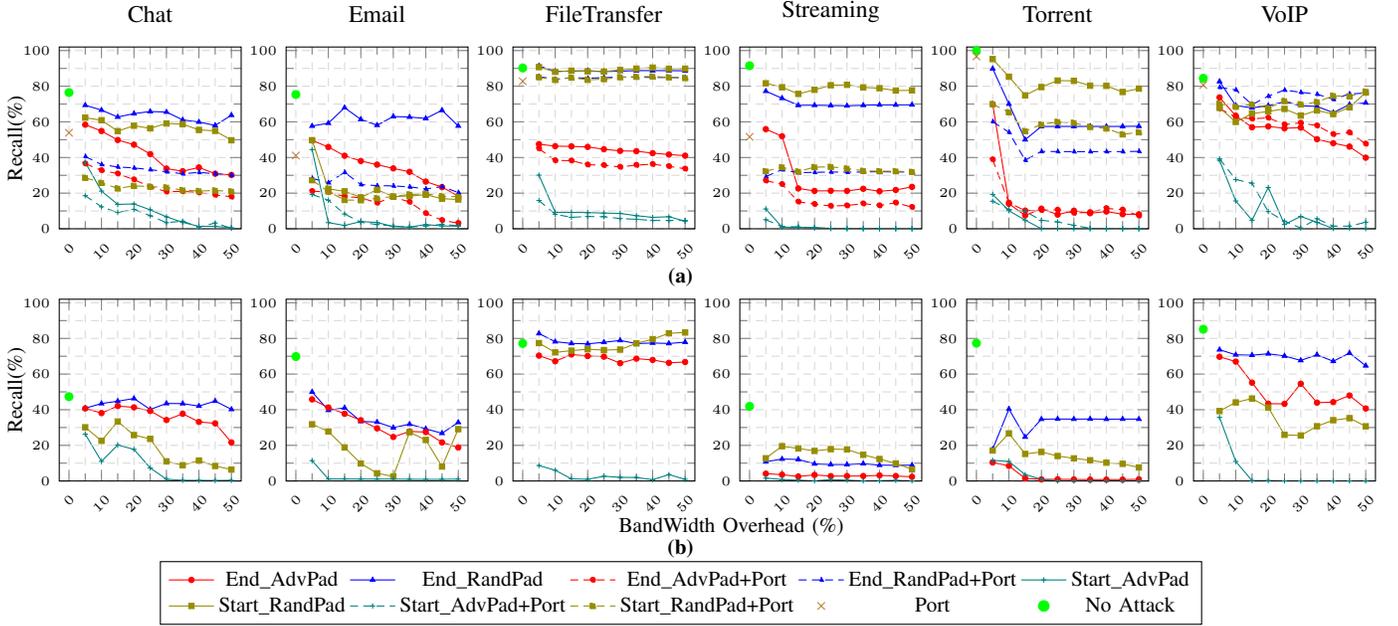

%% file: content/eval_flow_classification.tex
We evaluate the robustness of flow content classifiers against adversarial payload (AdvPay) attack. 
Also, we propose Random Payload (RandPay) attack as a baseline attack to show the effectiveness of the AdvPay attack. In the RandPay attack, a dummy packet that carries a payload with a random byte sequence is injected into the first n packets of a flow.
In the AdvPay attack, for a given class, Algorithm \ref{alg:AdversarialPayload} generates a UAP with the desired size on flows of that class in the validation set, then for each flow of that class in the test set, this perturbation is injected into the payload of a dummy packet, and the robustness of flow content classifiers is evaluated against them.

\subsubsection{Flow Content Classification Dataset}
Similar to packet classification, the decimal value of each byte is divided by 255 to normalize the byte sequence of packets between zero and one, and then, zero pad is added to the end of packets until their length reaches to $MaxPktSize$. Afterwards, the byte sequences of the first n packets of a flow which have at least one byte of payload are concatenated together, and this new sequence is called flow byte sequence. 
Table \ref{tab:FCC_Dataset} indicates the number of samples in each class before and after oversampling. 

\input{content/chart/fcc_dataset}

\subsubsection{Flow Content Classifiers Setup}
A 1D-CNN is used to classify the byte sequence of flows. Given our Computational resource, 150 1D-CNN with various architectures and hyperparameters have been explored to find the best classifier for FCC. Eventually, the classifier with higher accuracy on the validation set has been chosen. 
Based on experiments, we chose 10 packets to be in each flow byte sequence. 
The overall accuracy of FCC-HP and FCC-P is 81.52\% and 80.6\%, respectively. Other performance metrics of these classifiers are presented in Table \ref{tab:FCCs_performance_metrics}. The results demonstrate that the influence of the header of transport layer on FCC is low, and even in some classes, the performance of FCC-P is better than FCC-HP. 

\input{content/chart/fcc_eval}
\input{content/chart/fcc_perf}

\subsubsection{Flow Content Classifiers Robustness Evaluation}

AdvPay, RandPay, and port attack have been applied to flow content classifiers to evaluate their robustness. AdvPay attack has been run for 1000 iterations with $batch\_size=64$ and $\epsilon=0.001$. The dummy packet is injected after the first packet from source (\textit{e.g.}, client) to destination (\textit{e.g.}, server). RandPay attack has been conducted 50 times, and the average results are reported in this section.

Figure \ref{fig:fcc_evala} shows the recall of FCC-HP under AdvPay, RandPay, and port attack over various sizes of adversarial payload for all classes. 
The recall of FCC-HP under port attack is slightly reduced, which shows FCC-HP does not rely too much on the information of the port numbers of transport layer header. 
The recall of all classes under AdvPay attack drops to less than 3.52\% with just 500 bytes of adversarial payload. AdvPay+Port combines AdvPay with port attack. It increases the performance of the AdvPay attack a little bit and decreases the recall of all classes under AdvPay+Port to less than 1.41\% with just 500 bytes of adversarial payload.
The RandPay attack injects a dummy packet that carries a random byte sequence as the payload into a flow. Similar to AdvPay, the dummy packet is sent after the first packet from source to destination. Although the performance of RandPay is good, AdvPay attack has much better performance.
RandPay+Port combines RandPay with port attack. The performance of RandPay+Port is very close to RandPay. Unlike AdvPay, by increasing the size of adversarial payload, the recall of FCC-HP under RandPay and RandPay+Port remains relatively fixed. However, the recall of FCC-HP under AdvPay and AdvPay+Port is decreased by increasing the size of adversarial payload.

Figure \ref{fig:fcc_evalb} indicates the recall of FCC-P under AdvPay and RandPay over various sizes of adversarial payload for all classes. 
Except for the torrent class, the recall for all classes under AdvPay drops off to almost 0\% with 1000 or fewer bytes of adversarial payload. The recall of the torrent class decreases to 0\% with 1400 bytes adversarial payload. RandPad has poor performance on FCC-P in most classes, and by increasing the size of adversarial payload, the performance of RandPay does not improve very much. This observation demonstrates the effectiveness of the AdvPay attack on FCC-P. 
Although FCC-P has almost 1\% less overall accuracy than FCC-HP, results in Figure \ref{fig:fcc_eval} indicate FCC-P is slightly more robust than FCC-HP, and for reducing the performance of FCC-P, the size of adversarial payload needs to be increased more. 

%% file: content/chart/fcc_dataset.tex
\begin{table}[]
	\centering
	\caption{Flow content classification dataset.}
	\label{tab:FCC_Dataset}
	\setlength\extrarowheight{1pt}
	\resizebox{\linewidth}{!}{%
		\begin{tabular}{|
				>{\columncolor[HTML]{EFEFEF}}c |c|c|c|c|}
			\hline
			\cellcolor[HTML]{EFEFEF} & \multicolumn{2}{c|}{\cellcolor[HTML]{EFEFEF}Imbalanced Dataset} & \multicolumn{2}{c|}{\cellcolor[HTML]{EFEFEF}Balanced Dataset} \\ \cline{2-5} 
			\multirow{-2}{*}{\cellcolor[HTML]{EFEFEF}Type} & \cellcolor[HTML]{EFEFEF}Total Number & \cellcolor[HTML]{EFEFEF}Training Set & \cellcolor[HTML]{EFEFEF}Total Number & \cellcolor[HTML]{EFEFEF}Training Set \\ \hline
			Chat & 536 & 321 & 1174 & 959 \\ \hline
			Email & 392 & 235 & 1116 & 959 \\ \hline
			FileTransfer & 1420 & 852 & 1527 & 959 \\ \hline
			Streaming & 1114 & 668 & 1405 & 959 \\ \hline
			Torrent & 400 & 240 & 1119 & 959 \\ \hline
			VoIP & 1598 & 959 & 1598 & 959 \\ \hline
		\end{tabular}
	}
\end{table}

%% file: content/chart/fcc_eval.tex
\newcommand{\FCCchatr}{FCC_cls_chat_recall.csv}
\newcommand{\FCCemailr}{FCC_cls_email_recall.csv}
\newcommand{\FCCfiler}{FCC_cls_filetransfer_recall.csv}
\newcommand{\FCCstreamingr}{FCC_cls_streaming_recall.csv}
\newcommand{\FCCtorrentr}{FCC_cls_torrent_recall.csv}
\newcommand{\FCCvoipr}{FCC_cls_voip_recall.csv}
\newcommand{\FCCHP}{csv/FCC-HP/}
\newcommand{\FCCacccsv}{FCC_cls_Acc.csv}
\newcommand{\FCCP}{csv/FCC-P/}

\begin{figure*}[!t]
	\centering
	\begin{subfigure}{\linewidth}
		\centering
		\begin{tikzpicture}
		\pgfplotsset{
			scaled y ticks = false,
			width=4cm,
			height=4cm,
			axis on top,
			xmin=-50,xmax=1500,
			ymin=-0,ymax=100,
			minor y tick num=1,
			label style = {font = {\fontsize{8 pt}{12 pt}\selectfont}},
			title style = {font = {\fontsize{9pt}{12 pt}\selectfont}},
			tick label style = {font = {\fontsize{4 pt}{12 pt}\selectfont}},
			xtick={10,100,300,500,750,1000,1200,1400},
			x tick label style={rotate=45},
			ylabel shift={-0.8em},
			ylabel style={align=center},
			xlabel shift={-0.4em},
			grid=both, 
			grid style={dashed,gray!30}
		}
		\begin{groupplot}[ 
		group style={
			group size=6 by 1,
			vertical sep=55pt,
			horizontal sep=15pt
		},
		]
		\nextgroupplot[ xlabel={},ylabel={Recall(\%)},title={Chat},]
		\addplot[\colorone,mark=\markone,mark size=1.0]    table[x=AdvPaySize,y=rand,col sep=comma]{\FCCHP\FCCchatr};
		\addplot[\colortwo,mark=\marktwo,mark size=1.0]    table[x=AdvPaySize,y=advPay,col sep=comma]{\FCCHP\FCCchatr};  
		\addplot[\colorone,densely dashed ,mark=\markone,mark size=1.0]    table[x=AdvPaySize,y=rand_port,col sep=comma]{\FCCHP\FCCchatr};
		\addplot[\colortwo,densely dashed ,mark=\marktwo,mark size=1.0]    table[x=AdvPaySize,y=advPay_port,col sep=comma]{\FCCHP\FCCchatr};
		\addplot[only marks,\colorfive,mark=\markfive]    table[x=AdvPaySize,y=natural,col sep=comma]{\FCCHP\FCCchatr};  
		\addplot[only marks,\colorsix,mark=\marksix]    table[x=AdvPaySize,y=port,col sep=comma]{\FCCHP\FCCchatr};   
		\coordinate (c1) at (rel axis cs:0,1);
		
		\nextgroupplot[ xlabel={},ylabel={},title={Email}]
		\addplot[\colorone,mark=\markone,mark size=1.0]    table[x=AdvPaySize,y=rand,col sep=comma]{\FCCHP\FCCemailr};
		\addplot[\colortwo,mark=\marktwo,mark size=1.0]    table[x=AdvPaySize,y=advPay,col sep=comma]{\FCCHP\FCCemailr};  
		\addplot[\colorone,densely dashed ,mark=\markone,mark size=1.0]    table[x=AdvPaySize,y=rand_port,col sep=comma]{\FCCHP\FCCemailr};
		\addplot[\colortwo,densely dashed ,mark=\marktwo,mark size=1.0]    table[x=AdvPaySize,y=advPay_port,col sep=comma]{\FCCHP\FCCemailr};
		\addplot[only marks,\colorfive,mark=\markfive]    table[x=AdvPaySize,y=natural,col sep=comma]{\FCCHP\FCCemailr};  
		\addplot[only marks,\colorsix,mark=\marksix]    table[x=AdvPaySize,y=port,col sep=comma]{\FCCHP\FCCemailr};   
		
		\nextgroupplot[
		yshift=0pt,
		,x label style={at={(axis description cs:1.2,-0.25)},anchor=north},ylabel={},title={FileTransfer},
		legend to name=fred
		]
		\addplot[\colortwo,mark=\marktwo,mark size=1.0]    table[x=AdvPaySize,y=advPay,col sep=comma]{\FCCHP\FCCfiler};  
		\addplot[\colorone,mark=\markone,mark size=1.0]    table[x=AdvPaySize,y=rand,col sep=comma]{\FCCHP\FCCfiler};
		\addplot[\colortwo,densely dashed ,mark=\marktwo,mark size=1.0]    table[x=AdvPaySize,y=advPay_port,col sep=comma]{\FCCHP\FCCfiler};
		\addplot[\colorone,densely dashed ,mark=\markone,mark size=1.0]    table[x=AdvPaySize,y=rand_port,col sep=comma]{\FCCHP\FCCfiler};
		\addplot[only marks,\colorsix,mark=\marksix]    table[x=AdvPaySize,y=port,col sep=comma]{\FCCHP\FCCfiler};   
		\addplot[only marks,\colorfive,mark=\markfive]    table[x=AdvPaySize,y=natural,col sep=comma]{\FCCHP\FCCfiler};  
		
		
		\coordinate (c2) at (rel axis cs:1,1);
		\nextgroupplot[ xlabel={},ylabel={},title={Streaming},]
		\addplot[\colorone,mark=\markone,mark size=1.0]    table[x=AdvPaySize,y=rand,col sep=comma]{\FCCHP\FCCstreamingr};
		\addplot[\colortwo,mark=\marktwo,mark size=1.0]    table[x=AdvPaySize,y=advPay,col sep=comma]{\FCCHP\FCCstreamingr};  
		\addplot[\colorone,densely dashed ,mark=\markone,mark size=1.0]    table[x=AdvPaySize,y=rand_port,col sep=comma]{\FCCHP\FCCstreamingr};
		\addplot[\colortwo,densely dashed ,mark=\marktwo,mark size=1.0]    table[x=AdvPaySize,y=advPay_port,col sep=comma]{\FCCHP\FCCstreamingr};
		\addplot[only marks,\colorfive,mark=\markfive]    table[x=AdvPaySize,y=natural,col sep=comma]{\FCCHP\FCCstreamingr};  
		\addplot[only marks,\colorsix,mark=\marksix]    table[x=AdvPaySize,y=port,col sep=comma]{\FCCHP\FCCstreamingr};   
		\nextgroupplot[ xlabel={},ylabel={},title={Torrent},]
		\addplot[\colorone,mark=\markone,mark size=1.0]    table[x=AdvPaySize,y=rand,col sep=comma]{\FCCHP\FCCtorrentr};
		\addplot[\colortwo,mark=\marktwo,mark size=1.0]    table[x=AdvPaySize,y=advPay,col sep=comma]{\FCCHP\FCCtorrentr};  
		\addplot[\colorone,densely dashed ,mark=\markone,mark size=1.0]    table[x=AdvPaySize,y=rand_port,col sep=comma]{\FCCHP\FCCtorrentr};
		\addplot[\colortwo,densely dashed ,mark=\marktwo,mark size=1.0]    table[x=AdvPaySize,y=advPay_port,col sep=comma]{\FCCHP\FCCtorrentr};
		\addplot[only marks,\colorfive,mark=\markfive]    table[x=AdvPaySize,y=natural,col sep=comma]{\FCCHP\FCCtorrentr};  
		\addplot[only marks,\colorsix,mark=\marksix]    table[x=AdvPaySize,y=port,col sep=comma]{\FCCHP\FCCtorrentr};   
		\nextgroupplot[ xlabel={},ylabel={},title={VoIP},]
		\addplot[\colorone,mark=\markone,mark size=1.0]    table[x=AdvPaySize,y=rand,col sep=comma]{\FCCHP\FCCvoipr};
		\addplot[\colortwo,mark=\marktwo,mark size=1.0]    table[x=AdvPaySize,y=advPay,col sep=comma]{\FCCHP\FCCvoipr};  
		\addplot[\colorone,densely dashed ,mark=\markone,mark size=1.0]    table[x=AdvPaySize,y=rand_port,col sep=comma]{\FCCHP\FCCvoipr};
		\addplot[\colortwo,densely dashed ,mark=\marktwo,mark size=1.0]    table[x=AdvPaySize,y=advPay_port,col sep=comma]{\FCCHP\FCCvoipr};
		\addplot[only marks,\colorfive,mark=\markfive]    table[x=AdvPaySize,y=natural,col sep=comma]{\FCCHP\FCCvoipr};  
		\addplot[only marks,\colorsix,mark=\marksix]    table[x=AdvPaySize,y=port,col sep=comma]{\FCCHP\FCCvoipr};

		\end{groupplot}
		\coordinate (c3) at ($(c1)!.5!(c2)$);
		\node[below] at (c2 |- current bounding box.south)
		{\pgfplotslegendfromname{fred}};
		\end{tikzpicture}%
				\vspace{-1.8\baselineskip}

		\caption{ }
				\label{fig:fcc_evala}
	\end{subfigure}
	\begin{subfigure}{\linewidth}
		\centering
		\begin{tikzpicture}
		\pgfplotsset{
			scaled y ticks = false,
			width=4cm,
			height=4cm,
			axis on top,
			xmin=-50,xmax=1500,
			ymin=-0,ymax=100,
			minor y tick num=1,
			label style = {font = {\fontsize{8 pt}{12 pt}\selectfont}},
			title style = {font = {\fontsize{9pt}{12 pt}\selectfont}},
			tick label style = {font = {\fontsize{4 pt}{12 pt}\selectfont}},
			xtick={10,100,300,500,750,1000,1200,1400},
			ticklabel style = {font=\tiny},
			x tick label style={rotate=45},
			ylabel shift={-0.8em},
			ylabel style={align=center},
			xlabel shift={-0.4em},
			grid=both, 
			grid style={dashed,gray!30}
		}
		\begin{groupplot}[ 
		group style={
			group size=6 by 1,
			vertical sep=55pt,
			horizontal sep=15pt
		},
		]
		\nextgroupplot[ xlabel={},ylabel={Recall(\%)}]
		\addplot[\colorone,mark=\markone,mark size=1.0]    table[x=AdvPaySize,y=rand,col sep=comma]{\FCCP\FCCchatr};
		\addplot[\colortwo,mark=\marktwo,mark size=1.0]    table[x=AdvPaySize,y=advPay,col sep=comma]{\FCCP\FCCchatr};  
		\addplot[only marks,\colorfive,mark=\markfive]    table[x=AdvPaySize,y=natural,col sep=comma]{\FCCP\FCCchatr};  
		
		\coordinate (c1) at (rel axis cs:0,1);
		
		\nextgroupplot[ xlabel={},ylabel={}]
		\addplot[\colorone,mark=\markone,mark size=1.0]    table[x=AdvPaySize,y=rand,col sep=comma]{\FCCP\FCCemailr};
		\addplot[\colortwo,mark=\marktwo,mark size=1.0]    table[x=AdvPaySize,y=advPay,col sep=comma]{\FCCP\FCCemailr};  
		\addplot[only marks,\colorfive,mark=\markfive]    table[x=AdvPaySize,y=natural,col sep=comma]{\FCCP\FCCemailr};  
		
		\nextgroupplot[
		yshift=0pt,
		xlabel={Adversarial Payload  Size (byte)},x label style={at={(axis description cs:1.2,-0.25)},anchor=north},ylabel={}]
		legend to name=fred
		]
		\addplot[\colortwo,mark=\marktwo,mark size=1.0]    table[x=AdvPaySize,y=advPay,col sep=comma]{\FCCP\FCCfiler};  
		\addplot[\colorone,mark=\markone,mark size=1.0]    table[x=AdvPaySize,y=rand,col sep=comma]{\FCCP\FCCfiler};
		
		\addplot[only marks,\colorfive,mark=\markfive]    table[x=AdvPaySize,y=natural,col sep=comma]{\FCCP\FCCfiler};   
		
		\coordinate (c2) at (rel axis cs:1,1);
		\nextgroupplot[ xlabel={},ylabel={}]
		\addplot[\colorone,mark=\markone,mark size=1.0]    table[x=AdvPaySize,y=rand,col sep=comma]{\FCCP\FCCstreamingr};
		\addplot[\colortwo,mark=\marktwo,mark size=1.0]    table[x=AdvPaySize,y=advPay,col sep=comma]{\FCCP\FCCstreamingr};  
		\addplot[only marks,\colorfive,mark=\markfive]    table[x=AdvPaySize,y=natural,col sep=comma]{\FCCP\FCCstreamingr};  
		\nextgroupplot[ xlabel={},ylabel={}]
		\addplot[\colorone,mark=\markone,mark size=1.0]    table[x=AdvPaySize,y=rand,col sep=comma]{\FCCP\FCCtorrentr};
		\addplot[\colortwo,mark=\marktwo,mark size=1.0]    table[x=AdvPaySize,y=advPay,col sep=comma]{\FCCP\FCCtorrentr};  
		\addplot[only marks,\colorfive,mark=\markfive]    table[x=AdvPaySize,y=natural,col sep=comma]{\FCCP\FCCtorrentr};   
		\nextgroupplot[ xlabel={},ylabel={}]
		\addplot[\colorone,mark=\markone,mark size=1.0]    table[x=AdvPaySize,y=rand,col sep=comma]{\FCCP\FCCvoipr};
		\addplot[\colortwo,mark=\marktwo,mark size=1.0]    table[x=AdvPaySize,y=advPay,col sep=comma]{\FCCP\FCCvoipr};  
		\addplot[only marks,\colorfive,mark=\markfive]    table[x=AdvPaySize,y=natural,col sep=comma]{\FCCP\FCCvoipr};

		\end{groupplot}
		\coordinate (c3) at ($(c1)!.5!(c2)$);
		\node[below] at (c2 |- current bounding box.south)
		{\pgfplotslegendfromname{fred}};
		\end{tikzpicture}%
		
				\vspace{-0.9\baselineskip}

		\caption{ }
				\label{fig:fcc_evalb}
	\end{subfigure}
\begin{subfigure}{\linewidth}
	\centering
	\begin{tikzpicture} 
	\begin{axis}[%
	hide axis,
	xmin=10,
	xmax=50,
	ymin=0,
	ymax=0.4,
	legend style={nodes={scale=0.8, transform shape},at={($(0,0)+(1cm,1cm)$)},legend columns=-1,fill=none,draw=black,anchor=center,align=center}
	]
	\addlegendimage{\colortwo,mark=\marktwo,mark size=1.0}
	\addlegendentry{AdvPay};  
	\addlegendimage{\colorone,mark=\markone,mark size=1.0}
 	\addlegendentry{RandPay};     
	\addlegendimage{ \colortwo,densely dashed ,mark=\marktwo,mark size=1.0}
	\addlegendentry{AdvPay+Port};
	\addlegendimage{\colorone,densely dashed ,mark=\markone,mark size=1.0}
	\addlegendentry{RandPay+Port};  
	\addlegendimage{only marks,\colorsix,mark=\marksix}
    \addlegendentry{Port};      
	\addlegendimage{ only marks,\colorfive,mark=\markfive}
	\addlegendentry{No Attack};    
 
	\end{axis}
	\end{tikzpicture}
\end{subfigure}
	\caption{The recall of FCC-HP (a) and FCC-P (b) under different attacks over various sizes of adversarial payload for all classes. The legends show various kinds of attacks that have been applied.}
	\label{fig:fcc_eval}
\end{figure*}
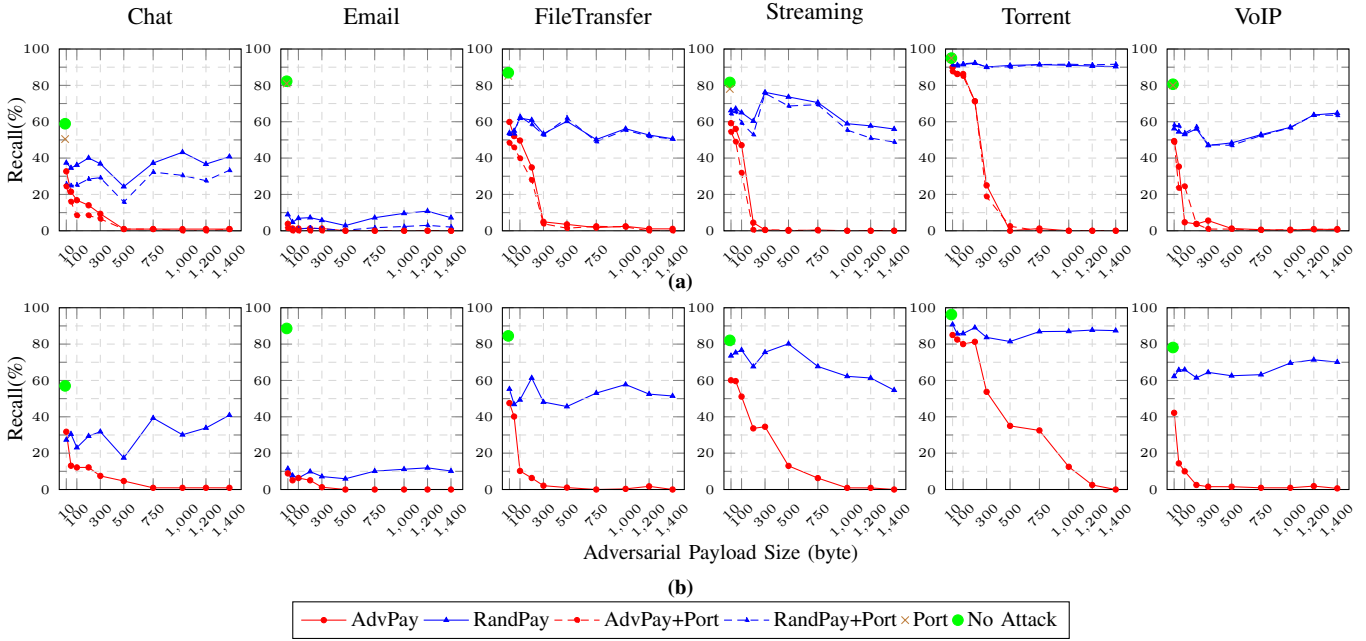

%% file: content/chart/fcc_perf.tex
\begin{table}[]
	\centering
	\setlength\extrarowheight{2pt}
	\caption{Performance metrics of the flow content classifiers FCC-HP, and FCC-P.}
	\label{tab:FCCs_performance_metrics}
	\resizebox{\linewidth}{!}{%
		\begin{tabular}{|
				>{\columncolor[HTML]{EFEFEF}}c |cc|cc|cc|}
			\hline
			\cellcolor[HTML]{EFEFEF} & \multicolumn{2}{c|}{\cellcolor[HTML]{EFEFEF}Precision(\%)} & \multicolumn{2}{c|}{\cellcolor[HTML]{EFEFEF}Recall(\%)} & \multicolumn{2}{c|}{\cellcolor[HTML]{EFEFEF}F-score(\%)} \\ \cline{2-7} 
			\multirow{-2}{*}{\cellcolor[HTML]{EFEFEF}Type} & \cellcolor[HTML]{EFEFEF}FCC-HP & \cellcolor[HTML]{EFEFEF}FCC-P & \cellcolor[HTML]{EFEFEF}FCC-HP & \cellcolor[HTML]{EFEFEF}FCC-P & \cellcolor[HTML]{EFEFEF}FCC-HP & \cellcolor[HTML]{EFEFEF}FCC-P \\ \hline
			Chat & 73.26 & 74.39 & 58.88 & 57.01 & 65.29 & 64.55 \\ \hline
			Email & 89.04 & 85.37 & 82.28 & 88.61 & 85.53 & 86.96 \\ \hline
			FileTransfer & 85.17 & 90.23 & 86.97 & 84.51 & 86.06 & 87.28 \\ \hline
			Streaming & 79.82 & 74.69 & 81.61 & 82.06 & 80.71 & 78.25 \\ \hline
			Torrent & 98.7 & 97.47 & 95.00 & 96.25 & 96.81 & 96.86 \\ \hline
			VoIP & 76.11 & 73.75 & 80.62 & 78.12 & 78.30 & 75.87 \\ \hline
		\end{tabular}
	}
\end{table}

%% file: content/eval-ts-classification.tex
We evaluate the robustness of flow time series classifiers against adversarial burst (AdvBurst) attack.  Also, we conduct RandBurst on these classifiers to evaluate the effectiveness of the AdvBurst attack. In the RandBurst attack, the statistical features of dummy packets are randomly perturbed. AdvBurst must have better performance than RandBurst. 
AdvBurst using Algorithm \ref{alg:AdversarialBurst} generates a UAP with a specific size on the flows of a given class in the validation set. Then, this perturbation is used as the statistical feature of dummy packets, which are appended to the end of selected bursts of flows of that class in the test set, and the robustness of classifiers is evaluated against these flows. 


\subsubsection{Flow Time Series Classification Dataset}
The number of samples in the dataset of FTSC is very close to the FCC dataset (Table \ref{tab:FCC_Dataset}). The packets size data are normalized as follows:
\footnotesize
\begin{equation}
\label{equ:standard_norm}
\begin{split}
&Normalized\; Packet Size = \frac{Packet Size - \mu_{PktSize}}{Std_{PktSize}}
\end{split}
\end{equation}
\footnotesize
where $\mu_{PktSize}$, and  $Std_{PktSize}$ indicate mean and standard deviation of packets size data, respectively. The IAT data are as follows:
\small
\begin{equation}
\label{equ:log_norm}
\begin{split}
&Normalized\; IAT = 2 \times \log_{IAT_{max}}^{(IAT + 1\mu Sec) }
\end{split}
\end{equation}
\normalsize
where $IAT_{max}$ is the maximum of IAT data.
\comment{The domain of packet size is between the minimum header size and $MaxPktSize$. The packets size is first normalized by standard normalization (Equation \ref{equ:standard_norm}), and later the minimum of normalized data is added to them, which makes normalized data positive. Finally, the domain of normalized data is limited between zero and one by dividing normalized data in the maximum of them. In the next step, the normalized packet size vector of each flow is multiplied to the vector of the direction sign (Equation \ref{eq:FTSC_data}).
\small
\begin{equation}
\label{equ:standard_norm}
\begin{split}
&Normalizzed\; Packet\; Size = \frac{Packet Size - \mu_{PktSize}}{Std_{PktSize}}
\end{split}
\end{equation}
\normalsize
where $\mu_{PktSize}$, and  $Std_{PktSize}$ indicate mean and standard deviation of packets size, respectively.
The domain of inter-arrival time is between zero microseconds and the longest flow timeout. The flow timeout is configurable, and in our system, the longest flow timeout is 180 seconds. The log normalization (Equation \ref{equ:log_norm}) is used to normalize inter-arrival time data, then the vector of the direction sign of packets is multiplied to them.
\small
\begin{equation}
\label{equ:log_norm}
\begin{split}
&Normalizzed\; IAT = 2 \times \log_{IAT_{max}}^{(IAT + 1\mu Sec) }
\end{split}
\end{equation}
\normalsize
where $IAT_{max}$ is the maximum of inter-arrival times data.
}
\subsubsection{Flow Time Series Classifiers Setup}
A 1D-CCN is used to classify flow time series. Given our computational resource, we have explored 150 different architectures and hyperparameters to build the final flow time series classifier. 
Based on experiments, we chose 100 packets to be in the time series of each flow. 
The overall accuracy of FTSC-PS and FTSC-IAT 
is 76.21\% and 76.51\%, respectively. The other performance metrics of these classifiers are reported on Table \ref{tab:FTSC-Per_Met}.  Based on our experiments, FTSC-IAT has a little bit better performance than FTSC-PS. 

\input{content/chart/ftsc_perf}

\input{content/chart/ftsc_eval}
\subsubsection{Flow Time Series Classifiers Robustness Evaluation}

AdvBurst and RandBurst have been applied to flow time series classifiers to evaluate their robustness. AdvBurst attack has been conducted for 2000 iterations with $batch\_siz=64$ and $\epsilon=0.01$. 
RandBurst has been conducted 50 times, and the average results are reported in this section.
The domain of IAT is between 0 and 180 seconds with microsecond granularity. If the IAT domain of dummy packets be equal to the domain of IAT in the dataset, AdvBurst can impose high time overhead on a flow. 
Therefore, we limit the IAT domain of dummy packets between 0.001 and 0.1 seconds, and this restriction is applied using the clip function in Algorithm \ref{alg:AdversarialBurst}.

We choose the first burst from source to destination as the selected burst in all flows to attack FTSC-PS. 
Figure \ref{fig:ftsc_evala} shows the recall  of FTSC-PS under AdvBurst and RandBurst over different numbers of dummy packets for all classes. The RandBurst attack appends dummy packets with random statistical features at the end of the selected bursts.
AdvBurst highly decreases the recall of FTSC-PS in all classes with just a few dummy packets. In chat, email, and torrent classes, the recall of FTSC-PS under AdvBurst drops to less than 11.65\% using just five dummy packets. Also, RandBurst shows excellent performance in these classes. Although AdvBurst has better performance on these classes, results show those classes are highly vulnerable to little manipulation in the size of dummy packets, and even dummy packets with random packet sizes can highly decrease the recall for these classes. In contrast, although the performance of AdvBurst in file transfer, streaming, and VoIP classes, is not as fine as those classes, the distance between the recall of FTSC-PS under AdvBurst and RandBurst is large. This observation indicates the effectiveness of AdvBurst in these classes.

We consider the first burst from destination to source in a flow as the selected burst
to attack on FTSC-IAT. Figure \ref{fig:ftsc_evalb} shows the recall of FTSC-IAT under AdvBurst and RandBurst for all classes. AdvBurst highly reduces the recall of FTSC-IAT for all classes with just a few dummy packets. However, its performance is various in different classes. The recall of FTSC-IAT under AdvBurst becomes less than 16.4\% for chat, torrent, and VoIP classes with just seven dummy packets.  AdvBurst needs more dummy packets to highly decrease the recall of FTSC-IAT for email and file transfer classes. 
Although the performance of AdvBurst on FTSC-IAT is fine enough, the gap between the recall of FTSC-IAT under AdvBurst and RandBurst is relatively small in most classes. 
We think the effectiveness of AdvBurst has decreased because of the domain of IAT has been limited between 0.001 and 0.1 seconds, and this domain is remarkably shorter than the real domain of IAT in the dataset. The recall of streaming class under AdvBurst has a strange behavior by increasing the number of dummy packets. The recall decreases continuously until the seventh dummy packet is added, and it begins to rise after that. We think the reason for this behavior is that the size of the burst heading from destination to source is often large in streaming class, and when the size of adversarial burst gets increased by increasing the number of dummy packets, FTSC-IAT recognizes these flows as streaming.

%% file: content/chart/ftsc_perf.tex
\begin{table}[]
	\centering
	\setlength\extrarowheight{2pt}
	\caption{Performance metrics of the flow time series classifiers FTSC-PS, and FTSC-IAT.}
	\label{tab:FTSC-Per_Met}
	\resizebox{\linewidth}{!}{%
		\begin{tabular}{|
				>{\columncolor[HTML]{EFEFEF}}c |cc|cc|cc|}
			\hline
			\cellcolor[HTML]{EFEFEF} & \multicolumn{2}{c|}{\cellcolor[HTML]{EFEFEF}Precision(\%)} & \multicolumn{2}{c|}{\cellcolor[HTML]{EFEFEF}Recall(\%)} & \multicolumn{2}{c|}{\cellcolor[HTML]{EFEFEF}F-score(\%)} \\ \cline{2-7} 
			\multirow{-2}{*}{\cellcolor[HTML]{EFEFEF}Type} & \cellcolor[HTML]{EFEFEF}{\scriptsize FTSC-PS} & \cellcolor[HTML]{EFEFEF}{\scriptsize FTSC-IAT}& \cellcolor[HTML]{EFEFEF}{\scriptsize FTSC-PS} & \cellcolor[HTML]{EFEFEF}{\scriptsize FTSC-IAT} & \cellcolor[HTML]{EFEFEF}{\scriptsize FTSC-PS} & \cellcolor[HTML]{EFEFEF}{\scriptsize FTSC-IAT} \\ \hline
			Chat & 58.25 & 65.22 & 58.25 & 58.25 & 58.25 & 61.54 \\ \hline
			Email & 92.21 & 85.54 & 89.87 & 89.87 & 91.02 & 87.65 \\ \hline
			FileTransfer & 80.00 & 80.95 & 76.60 & 78.37 & 78.26 & 79.64 \\ \hline
			Streaming & 78.28 & 75.00 & 71.10 & 70.18 & 74.52 & 72.51 \\ \hline
			Torrent & 88.24 & 90.24 & 93.75 & 92.50 & 90.91 & 91.36 \\ \hline
			VoIP & 70.85 & 71.39 & 77.39 & 77.81 & 73.98 & 74.46 \\ \hline
		\end{tabular}
	}
\end{table}

%% file: content/chart/ftsc_eval.tex
\newcommand{\FTSCchatr}{FTSC_cls_chat_recall.csv}
\newcommand{\FTSCemailr}{FTSC_cls_email_recall.csv}
\newcommand{\FTSCfiler}{FTSC_cls_filetransfer_recall.csv}
\newcommand{\FTSCstreamingr}{FTSC_cls_streaming_recall.csv}
\newcommand{\FTSCtorrentr}{FTSC_cls_torrent_recall.csv}
\newcommand{\FTSCvoipr}{FTSC_cls_voip_recall.csv}
\newcommand{\FTSCPS}{csv/FTSC-PS/}
\newcommand{\FTSCacccsv}{FTSC_cls_Acc.csv}
\newcommand{\FTSCIAT}{csv/FTSC-IAT/}

\begin{figure*}[!t]
	\centering
	\begin{subfigure}{\linewidth}
		\centering
		\begin{tikzpicture}
		\pgfplotsset{
			scaled y ticks = false,
			width=4cm,
			height=4cm,
			axis on top,
			xmin=-1,xmax=21,
			ymin=-0,ymax=100,
			minor y tick num=1,
			xtick={1,3,5,7,10,12,15,17,20},
			label style = {font = {\fontsize{8 pt}{12 pt}\selectfont}},
			title style = {font = {\fontsize{9pt}{12 pt}\selectfont}},
			tick label style = {font = {\fontsize{4 pt}{12 pt}\selectfont}},
			x tick label style={rotate=45},
			ylabel shift={-0.8em},
			ylabel style={align=center},
			xlabel shift={-0.4em},
			grid=both, 
			grid style={dashed,gray!30}
		}
		\begin{groupplot}[ 
		group style={
			group size=6 by 1,
			vertical sep=45pt,
			horizontal sep=15pt
		},
		]
		\nextgroupplot[ xlabel={},ylabel={Recall(\%)},title={Chat},]
		\addplot[\colorone,mark=\markone,mark size=1.0]    table[x=AdvBurstOverHead,y=rand,col sep=comma]{\FTSCPS\FTSCchatr};
		\addplot[\colortwo,mark=\marktwo,mark size=1.0]    table[x=AdvBurstOverHead,y=AdvBurst,col sep=comma]{\FTSCPS\FTSCchatr};  
		\addplot[blue,dashed]    table[x=AdvBurstOverHead,y=rand_port,col sep=comma]{\FTSCPS\FTSCchatr};
		\addplot[red,dashed]    table[x=AdvBurstOverHead,y=AdvBurst_port,col sep=comma]{\FTSCPS\FTSCchatr};
		\addplot[only marks,\colorfive,mark=\markfive]    table[x=AdvBurstOverHead,y=natural,col sep=comma]{\FTSCPS\FTSCchatr};  
		\addplot[only marks,olive,mark=*]    table[x=AdvBurstOverHead,y=port,col sep=comma]{\FTSCPS\FTSCchatr};   
		\coordinate (c1) at (rel axis cs:0,1);
		
		\nextgroupplot[ xlabel={},ylabel={},title={Email}]
		\addplot[\colorone,mark=\markone,mark size=1.0]    table[x=AdvBurstOverHead,y=rand,col sep=comma]{\FTSCPS\FTSCemailr};
		\addplot[\colortwo,mark=\marktwo,mark size=1.0]    table[x=AdvBurstOverHead,y=AdvBurst,col sep=comma]{\FTSCPS\FTSCemailr};  
		\addplot[blue,dashed]    table[x=AdvBurstOverHead,y=rand_port,col sep=comma]{\FTSCPS\FTSCemailr};
		\addplot[red,dashed]    table[x=AdvBurstOverHead,y=AdvBurst_port,col sep=comma]{\FTSCPS\FTSCemailr};
		\addplot[only marks,\colorfive,mark=\markfive]    table[x=AdvBurstOverHead,y=natural,col sep=comma]{\FTSCPS\FTSCemailr};  
		\addplot[only marks,olive,mark=*]    table[x=AdvBurstOverHead,y=port,col sep=comma]{\FTSCPS\FTSCemailr};   
		
		\nextgroupplot[
		,ylabel={},title={FileTransfer},
		]
		\addplot[\colortwo,mark=\marktwo,mark size=1.0]    table[x=AdvBurstOverHead,y=AdvBurst,col sep=comma]{\FTSCPS\FTSCfiler};  
		\addplot[\colorone,mark=\markone,mark size=1.0]    table[x=AdvBurstOverHead,y=rand,col sep=comma]{\FTSCPS\FTSCfiler};
		\addplot[only marks,\colorfive,mark=\markfive]    table[x=AdvBurstOverHead,y=natural,col sep=comma]{\FTSCPS\FTSCfiler};  
		 
		\coordinate (c2) at (rel axis cs:1,1);
		\nextgroupplot[ xlabel={},ylabel={},title={Streaming},]
		\addplot[\colorone,mark=\markone,mark size=1.0]    table[x=AdvBurstOverHead,y=rand,col sep=comma]{\FTSCPS\FTSCstreamingr};
		\addplot[\colortwo,mark=\marktwo,mark size=1.0]    table[x=AdvBurstOverHead,y=AdvBurst,col sep=comma]{\FTSCPS\FTSCstreamingr};  
		\addplot[blue,dashed]    table[x=AdvBurstOverHead,y=rand_port,col sep=comma]{\FTSCPS\FTSCstreamingr};
		\addplot[red,dashed]    table[x=AdvBurstOverHead,y=AdvBurst_port,col sep=comma]{\FTSCPS\FTSCstreamingr};
		\addplot[only marks,\colorfive,mark=\markfive]    table[x=AdvBurstOverHead,y=natural,col sep=comma]{\FTSCPS\FTSCstreamingr};  
		\addplot[only marks,olive,mark=*]    table[x=AdvBurstOverHead,y=port,col sep=comma]{\FTSCPS\FTSCstreamingr};   
		\nextgroupplot[ xlabel={},ylabel={},title={Torrent},]
		\addplot[\colorone,mark=\markone,mark size=1.0]    table[x=AdvBurstOverHead,y=rand,col sep=comma]{\FTSCPS\FTSCtorrentr};
		\addplot[\colortwo,mark=\marktwo,mark size=1.0]    table[x=AdvBurstOverHead,y=AdvBurst,col sep=comma]{\FTSCPS\FTSCtorrentr};  
		\addplot[blue,dashed]    table[x=AdvBurstOverHead,y=rand_port,col sep=comma]{\FTSCPS\FTSCtorrentr};
		\addplot[red,dashed]    table[x=AdvBurstOverHead,y=AdvBurst_port,col sep=comma]{\FTSCPS\FTSCtorrentr};
		\addplot[only marks,\colorfive,mark=\markfive]    table[x=AdvBurstOverHead,y=natural,col sep=comma]{\FTSCPS\FTSCtorrentr};  
		\addplot[only marks,olive,mark=*]    table[x=AdvBurstOverHead,y=port,col sep=comma]{\FTSCPS\FTSCtorrentr};   
		\nextgroupplot[ xlabel={},ylabel={},title={VoIP},]
		\addplot[\colorone,mark=\markone,mark size=1.0]    table[x=AdvBurstOverHead,y=rand,col sep=comma]{\FTSCPS\FTSCvoipr};
		\addplot[\colortwo,mark=\marktwo,mark size=1.0]    table[x=AdvBurstOverHead,y=AdvBurst,col sep=comma]{\FTSCPS\FTSCvoipr};  
		\addplot[blue,dashed]    table[x=AdvBurstOverHead,y=rand_port,col sep=comma]{\FTSCPS\FTSCvoipr};
		\addplot[red,dashed]    table[x=AdvBurstOverHead,y=AdvBurst_port,col sep=comma]{\FTSCPS\FTSCvoipr};
		\addplot[only marks,\colorfive,mark=\markfive]    table[x=AdvBurstOverHead,y=natural,col sep=comma]{\FTSCPS\FTSCvoipr};  
		\addplot[only marks,olive,mark=*]    table[x=AdvBurstOverHead,y=port,col sep=comma]{\FTSCPS\FTSCvoipr};

		\end{groupplot}
		\coordinate (c3) at ($(c1)!.5!(c2)$);
		\node[below] at (c2 |- current bounding box.south)
		{\pgfplotslegendfromname{fred}};
		\end{tikzpicture}%
				\vspace{-1.5\baselineskip}

		\caption{ }
					\label{fig:ftsc_evala}
	\end{subfigure}
	\begin{subfigure}{\linewidth}
		\centering
		\begin{tikzpicture}
		\pgfplotsset{
			scaled y ticks = false,
			width=4cm,
			height=4cm,
			axis on top,
			label style = {font = {\fontsize{8 pt}{12 pt}\selectfont}},
			title style = {font = {\fontsize{9pt}{12 pt}\selectfont}},
			tick label style = {font = {\fontsize{4 pt}{12 pt}\selectfont}},
			x tick label style={rotate=45},
			xmin=-1,xmax=21,
			ymin=-0,ymax=100,
			minor y tick num=1,
			xtick={1,3,5,7,10,12,15,17,20},
			ylabel shift={-0.8em},
			ylabel style={align=center},
			xlabel shift={-0.4em},
			grid=both, 
			grid style={dashed,gray!30}
		}
		\begin{groupplot}[ 
		group style={
			group size=6 by 1,
			vertical sep=45pt,
			horizontal sep=15pt
		},
		]
		\nextgroupplot[ xlabel={},ylabel={Recall(\%)}]
		\addplot[\colorone,mark=\markone,mark size=1.0]    table[x=AdvBurstOverHead,y=rand,col sep=comma]{\FTSCIAT\FTSCchatr};
		\addplot[\colortwo,mark=\marktwo,mark size=1.0]    table[x=AdvBurstOverHead,y=AdvBurst,col sep=comma]{\FTSCIAT\FTSCchatr};  
		\addplot[blue,dashed]    table[x=AdvBurstOverHead,y=rand_port,col sep=comma]{\FTSCIAT\FTSCchatr};
		\addplot[red,dashed]    table[x=AdvBurstOverHead,y=AdvBurst_port,col sep=comma]{\FTSCIAT\FTSCchatr};
		\addplot[only marks,\colorfive,mark=\markfive]    table[x=AdvBurstOverHead,y=natural,col sep=comma]{\FTSCIAT\FTSCchatr};  
		\addplot[only marks,olive,mark=*]    table[x=AdvBurstOverHead,y=port,col sep=comma]{\FTSCIAT\FTSCchatr};   
		\coordinate (c1) at (rel axis cs:0,1);
		
		\nextgroupplot[ xlabel={},ylabel={}]
		\addplot[\colorone,mark=\markone,mark size=1.0]    table[x=AdvBurstOverHead,y=rand,col sep=comma]{\FTSCIAT\FTSCemailr};
		\addplot[\colortwo,mark=\marktwo,mark size=1.0]    table[x=AdvBurstOverHead,y=AdvBurst,col sep=comma]{\FTSCIAT\FTSCemailr};  
		\addplot[blue,dashed]    table[x=AdvBurstOverHead,y=rand_port,col sep=comma]{\FTSCIAT\FTSCemailr};
		\addplot[red,dashed]    table[x=AdvBurstOverHead,y=AdvBurst_port,col sep=comma]{\FTSCIAT\FTSCemailr};
		\addplot[only marks,\colorfive,mark=\markfive]    table[x=AdvBurstOverHead,y=natural,col sep=comma]{\FTSCIAT\FTSCemailr};  
		\addplot[only marks,olive,mark=*]    table[x=AdvBurstOverHead,y=port,col sep=comma]{\FTSCIAT\FTSCemailr};   
		
		\nextgroupplot[
		yshift=0pt,
		xlabel={Number of Dummy Packets},x label style={at={(axis description cs:1.2,-0.15)},anchor=north},ylabel={}
		]
		
		\addplot[\colortwo,mark=\marktwo,mark size=1.0]    table[x=AdvBurstOverHead,y=AdvBurst,col sep=comma]{\FTSCIAT\FTSCfiler};  
		\addplot[\colorone,mark=\markone,mark size=1.0]    table[x=AdvBurstOverHead,y=rand,col sep=comma]{\FTSCIAT\FTSCfiler};
		
		\addplot[only marks,\colorfive,mark=\markfive]    table[x=AdvBurstOverHead,y=natural,col sep=comma]{\FTSCIAT\FTSCfiler};  
		
		
		\coordinate (c2) at (rel axis cs:1,1);
		\nextgroupplot[ xlabel={},ylabel={}]
		\addplot[\colorone,mark=\markone,mark size=1.0]    table[x=AdvBurstOverHead,y=rand,col sep=comma]{\FTSCIAT\FTSCstreamingr};
		\addplot[\colortwo,mark=\marktwo,mark size=1.0]    table[x=AdvBurstOverHead,y=AdvBurst,col sep=comma]{\FTSCIAT\FTSCstreamingr};  
		\addplot[blue,dashed]    table[x=AdvBurstOverHead,y=rand_port,col sep=comma]{\FTSCIAT\FTSCstreamingr};
		\addplot[red,dashed]    table[x=AdvBurstOverHead,y=AdvBurst_port,col sep=comma]{\FTSCIAT\FTSCstreamingr};
		\addplot[only marks,\colorfive,mark=\markfive]    table[x=AdvBurstOverHead,y=natural,col sep=comma]{\FTSCIAT\FTSCstreamingr};  
		\addplot[only marks,olive,mark=*]    table[x=AdvBurstOverHead,y=port,col sep=comma]{\FTSCIAT\FTSCstreamingr};   
		\nextgroupplot[ xlabel={},ylabel={}]
		\addplot[\colorone,mark=\markone,mark size=1.0]    table[x=AdvBurstOverHead,y=rand,col sep=comma]{\FTSCIAT\FTSCtorrentr};
		\addplot[\colortwo,mark=\marktwo,mark size=1.0]    table[x=AdvBurstOverHead,y=AdvBurst,col sep=comma]{\FTSCIAT\FTSCtorrentr};  
		\addplot[blue,dashed]    table[x=AdvBurstOverHead,y=rand_port,col sep=comma]{\FTSCIAT\FTSCtorrentr};
		\addplot[red,dashed]    table[x=AdvBurstOverHead,y=AdvBurst_port,col sep=comma]{\FTSCIAT\FTSCtorrentr};
		\addplot[only marks,\colorfive,mark=\markfive]    table[x=AdvBurstOverHead,y=natural,col sep=comma]{\FTSCIAT\FTSCtorrentr};  
		\addplot[only marks,olive,mark=*]    table[x=AdvBurstOverHead,y=port,col sep=comma]{\FTSCIAT\FTSCtorrentr};   
		\nextgroupplot[ xlabel={},ylabel={}]
		\addplot[\colorone,mark=\markone,mark size=1.0]    table[x=AdvBurstOverHead,y=rand,col sep=comma]{\FTSCIAT\FTSCvoipr};
		\addplot[\colortwo,mark=\marktwo,mark size=1.0]    table[x=AdvBurstOverHead,y=AdvBurst,col sep=comma]{\FTSCIAT\FTSCvoipr};  
		\addplot[only marks,\colorfive,mark=\markfive]    table[x=AdvBurstOverHead,y=natural,col sep=comma]{\FTSCIAT\FTSCvoipr};  
		\addplot[only marks,olive,mark=*]    table[x=AdvBurstOverHead,y=port,col sep=comma]{\FTSCIAT\FTSCvoipr};

		\end{groupplot}
		\coordinate (c3) at ($(c1)!.5!(c2)$);
		\node[below] at (c2 |- current bounding box.south)
		{\pgfplotslegendfromname{fred}};
		\end{tikzpicture}%
				\vspace{-1.1\baselineskip}
	
		\caption{ }
			\label{fig:ftsc_evalb}
	\end{subfigure}
\begin{subfigure}{\linewidth}
	\centering
	\begin{tikzpicture} 
	\begin{axis}[%
	hide axis,
	xmin=10,
	xmax=50,
	ymin=0,
	ymax=0.4,
	legend style={nodes={scale=0.8, transform shape},at={($(0,0)+(1cm,1cm)$)},legend columns=-1,fill=none,draw=black,anchor=center,align=center}
	]
	\addlegendimage{\colortwo,mark=\marktwo,mark size=1.0}
	\addlegendentry{AdvBurst};  
	\addlegendimage{\colorone,mark=\markone,mark size=1.0}
	\addlegendentry{RandBurst};     
	\addlegendimage{ only marks,\colorfive,mark=\markfive}
	\addlegendentry{No Attack};    
	
	\end{axis}
	\end{tikzpicture}
\end{subfigure}
	\caption{The recall of FTSC-PS (a) and FTSC-IAT (b) under different attacks over various number of dummy packets for all classes. The legends show various kinds of attacks that have been applied.}
	\label{fig:ftsc_eval}
\end{figure*}
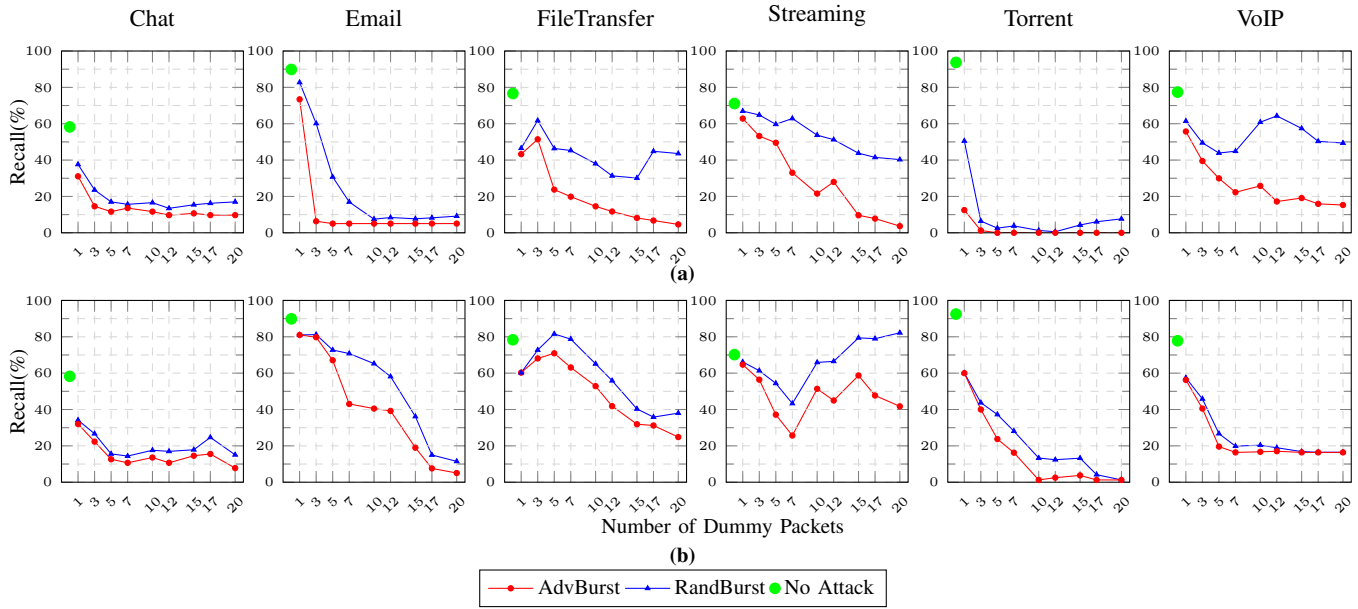

%% file: content/blackbox.tex
\begin{table*}[]
	\centering
	\setlength\extrarowheight{0.8pt}
	\caption{The recall and the overall accuracy of six new target classifiers against ANT. The target classifiers are Stacked Auto-Encoders (SAEs), and ANT is created using the 1D-CNNs classifiers in previous experiments.}
	\label{tab:bb}
	\resizebox{\linewidth}{!}{%
		\begin{tabular}{cllclclcccccc}
			& \multicolumn{1}{c}{}                         &                          &                                    &                          &                                        &                          & \multicolumn{6}{c}{Recall(\%)}                                                                                                                                                                \\ \cline{8-13} 
			& \multicolumn{1}{c}{\multirow{-2}{*}{Attack}} &                          & \multirow{-2}{*}{Attack Parameter} &                          & \multirow{-2}{*}{Overall Accuracy(\%)} &                          & Chat                          & Email                         & File Transfer                 & Streaming                     & Torrent                       & VoIP                          \\ \cline{2-2} \cline{4-4} \cline{6-6} \cline{8-13} 
			\multicolumn{1}{l}{}                            &                                              &                          &                                    &                          & \multicolumn{1}{l}{}                   &                          & \multicolumn{1}{l}{}          & \multicolumn{1}{l}{}          & \multicolumn{1}{l}{}          & \multicolumn{1}{l}{}          & \multicolumn{1}{l}{}          & \multicolumn{1}{l}{}          \\[-1ex]
			\multicolumn{1}{c|}{}                           & \cellcolor[HTML]{EFEFEF}No Attack            & \cellcolor[HTML]{EFEFEF} & \cellcolor[HTML]{EFEFEF}           & \cellcolor[HTML]{EFEFEF} & \cellcolor[HTML]{EFEFEF}82.01          & \cellcolor[HTML]{EFEFEF} & \cellcolor[HTML]{EFEFEF}67.83 & \cellcolor[HTML]{EFEFEF}79.04 & \cellcolor[HTML]{EFEFEF}90.52 & \cellcolor[HTML]{EFEFEF}80.03 & \cellcolor[HTML]{EFEFEF}96.00   & \cellcolor[HTML]{EFEFEF}74.04 \\
			\multicolumn{1}{c|}{}                           & End\_AdvPad                                  &                          & 20\% BWO                           &                          & 47.95                                  &                          & 42.75                         & 75.91                         & 71.44                         & 63.11                         & 0.65                          & 51.87                         \\
			\multicolumn{1}{c|}{}                           & End\_AdvPad + Port                           &                          & 20\% BWO                           &                          & 41.52                                  &                          & 19.67                         & 57.90                          & 69.63                         & 46.13                         & 0.27                          & 62.84                         \\
			\multicolumn{1}{c|}{}                           & Start\_AdvPad                                &                          & 20\% BWO                           &                          & 42.82                                  &                          & 30.66                         & 13.05                         & 79.27                         & 77.19                         & 0.61                          & 34.81                         \\
			\multicolumn{1}{c|}{\multirow{-5}{*}{PC-HP}}    & Start\_AdvPad + Port                         &                          & 20\% BWO                           &                          & 35.46                                  &                          & 22.81                         & 9.22                          & 68.67                         & 41.95                         & 0.16                          & 50.69                         \\
			\multicolumn{1}{l}{}                            &                                              &                          &                                    &                          & \multicolumn{1}{l}{}                   &                          & \multicolumn{1}{l}{}          & \multicolumn{1}{l}{}          & \multicolumn{1}{l}{}          & \multicolumn{1}{l}{}          & \multicolumn{1}{l}{}          & \multicolumn{1}{l}{}          \\[-1ex]
			\multicolumn{1}{c|}{}                           & \cellcolor[HTML]{EFEFEF}No Attack            & \cellcolor[HTML]{EFEFEF} & \cellcolor[HTML]{EFEFEF}           & \cellcolor[HTML]{EFEFEF} & \cellcolor[HTML]{EFEFEF}63.22          & \cellcolor[HTML]{EFEFEF} & \cellcolor[HTML]{EFEFEF}35.37 & \cellcolor[HTML]{EFEFEF}80.23 & \cellcolor[HTML]{EFEFEF}76.11 & \cellcolor[HTML]{EFEFEF}21.88 & \cellcolor[HTML]{EFEFEF}87.67 & \cellcolor[HTML]{EFEFEF}84.08 \\
			\multicolumn{1}{c|}{}                           & End\_AdvPad                                  &                          & 20\% BWO                           &                          & 36.15                                  &                          & 31.64                         & 68.70                          & 76.62                         & 9.29                          & 0.36                          & 51.33                         \\
			\multicolumn{1}{c|}{\multirow{-3}{*}{PC-P}}     & Start\_AdvPad                                &                          & 20\% BWO                           &                          & 8.42                                   &                          & 21.09                         & 2.91                          & 12.63                         & 10.20                          & 0.23                          & 2.19                          \\
			\multicolumn{1}{l}{}                            &                                              &                          &                                    &                          & \multicolumn{1}{l}{}                   &                          & \multicolumn{1}{l}{}          & \multicolumn{1}{l}{}          & \multicolumn{1}{l}{}          & \multicolumn{1}{l}{}          & \multicolumn{1}{l}{}          & \multicolumn{1}{l}{}          \\[-1ex]
			\multicolumn{1}{c|}{}                           & \cellcolor[HTML]{EFEFEF}No Attack            & \cellcolor[HTML]{EFEFEF} & \cellcolor[HTML]{EFEFEF}           & \cellcolor[HTML]{EFEFEF} & \cellcolor[HTML]{EFEFEF}77.40           & \cellcolor[HTML]{EFEFEF} & \cellcolor[HTML]{EFEFEF}50.47 & \cellcolor[HTML]{EFEFEF}86.08 & \cellcolor[HTML]{EFEFEF}80.63 & \cellcolor[HTML]{EFEFEF}78.03 & \cellcolor[HTML]{EFEFEF}95.00    & \cellcolor[HTML]{EFEFEF}76.56 \\
			\multicolumn{1}{c|}{}                           & AdvPay                                       &                          & 750 bytes                          &                          & 42.91                                  &                          & 31.78                         & 3.80                           & 42.61                         & 44.39                         & 81.25                         & 45.94                         \\
			\multicolumn{1}{c|}{\multirow{-3}{*}{FCC-HP}}   & AdvPay + Port                                &                          & 750 bytes                          &                          & 41.35                                  &                          & 23.36                         & 1.27                          & 34.51                         & 43.95                         & 82.50                          & 51.25                         \\
			\multicolumn{1}{l}{}                            &                                              &                          &                                    &                          & \multicolumn{1}{l}{}                   &                          & \multicolumn{1}{l}{}          & \multicolumn{1}{l}{}          & \multicolumn{1}{l}{}          & \multicolumn{1}{l}{}          & \multicolumn{1}{l}{}          & \multicolumn{1}{l}{}          \\[-1ex]
			\multicolumn{1}{c|}{}                           & \cellcolor[HTML]{EFEFEF}No Attack            & \cellcolor[HTML]{EFEFEF} & \cellcolor[HTML]{EFEFEF}           & \cellcolor[HTML]{EFEFEF} & \cellcolor[HTML]{EFEFEF}76.85          & \cellcolor[HTML]{EFEFEF} & \cellcolor[HTML]{EFEFEF}47.66 & \cellcolor[HTML]{EFEFEF}84.81 & \cellcolor[HTML]{EFEFEF}80.99 & \cellcolor[HTML]{EFEFEF}78.03 & \cellcolor[HTML]{EFEFEF}95.00   & \cellcolor[HTML]{EFEFEF}75.62 \\
			\multicolumn{1}{c|}{\multirow{-2}{*}{FCC-P}}    & AdvPay                                       &                          & 750 bytes                          &                          & 43.18                                  &                          & 23.36                         & 6.33                          & 36.27                         & 52.47                         & 81.25                         & 49.06                         \\
			\multicolumn{1}{l}{}                            &                                              &                          &                                    &                          & \multicolumn{1}{l}{}                   &                          & \multicolumn{1}{l}{}          & \multicolumn{1}{l}{}          & \multicolumn{1}{l}{}          & \multicolumn{1}{l}{}          & \multicolumn{1}{l}{}          & \multicolumn{1}{l}{}          \\[-1ex]
			\multicolumn{1}{c|}{}                           & \cellcolor[HTML]{EFEFEF}No Attack            & \cellcolor[HTML]{EFEFEF} & \cellcolor[HTML]{EFEFEF}           & \cellcolor[HTML]{EFEFEF} & \cellcolor[HTML]{EFEFEF}69.62          & \cellcolor[HTML]{EFEFEF} & \cellcolor[HTML]{EFEFEF}34.95 & \cellcolor[HTML]{EFEFEF}75.95 & \cellcolor[HTML]{EFEFEF}73.76 & \cellcolor[HTML]{EFEFEF}66.51 & \cellcolor[HTML]{EFEFEF}85.00    & \cellcolor[HTML]{EFEFEF}73.95 \\
			\multicolumn{1}{c|}{\multirow{-2}{*}{FTSC-IAT}} & AdvBurst                                     &                          & 7 dummy pkts                       &                          & 31.31                                  &                          & 5.83                          & 37.97                         & 19.50                          & 81.65                         & 3.75                          & 20.58                         \\
			\multicolumn{1}{l}{}                            &                                              &                          &                                    &                          & \multicolumn{1}{l}{}                   &                          & \multicolumn{1}{l}{}          & \multicolumn{1}{l}{}          & \multicolumn{1}{l}{}          & \multicolumn{1}{l}{}          & \multicolumn{1}{l}{}          & \multicolumn{1}{l}{}          \\[-1ex]
			\multicolumn{1}{c|}{}                           & \cellcolor[HTML]{EFEFEF}No Attack            & \cellcolor[HTML]{EFEFEF} & \cellcolor[HTML]{EFEFEF}           & \cellcolor[HTML]{EFEFEF} & \cellcolor[HTML]{EFEFEF}67.94          & \cellcolor[HTML]{EFEFEF} & \cellcolor[HTML]{EFEFEF}48.54 & \cellcolor[HTML]{EFEFEF}79.75 & \cellcolor[HTML]{EFEFEF}67.38 & \cellcolor[HTML]{EFEFEF}62.84 & \cellcolor[HTML]{EFEFEF}88.75 & \cellcolor[HTML]{EFEFEF}70.06 \\
			\multicolumn{1}{c|}{\multirow{-2}{*}{FTSC-PS}}  & AdvBurst                                     &                          & 7 dummy pkts                       &                          & 30.76                                  &                          & 7.77                          & 6.33                          & 78.01                         & 22.94                         & 0.00                             & 15.29                        
		\end{tabular}
	}
\end{table*}

Adversarial examples are transferable among various classifiers and training sets. Since the limitation of the dataset, we only evaluate the transferability of ANT among various classifiers. We train six new Stacked Autoencoder (SAE) classifiers as the new target classifiers and use ANT that has been generated in previous experiments to evaluate the transferability of ANT. 
We utilize the SAE architecture of \cite{DBLP:journals/access/WangYCQ18} for the new packet and flow content classifiers and \cite{DBLP:conf/ndss/RimmerPJGJ18} for the new flow time series classifiers. In this experiment, there is no knowledge about the parameters of the target classifiers in the generation process of ANT. 
Table \ref{tab:bb} indicates the performance of the new target classifiers against ANT. The results demonstrate that ANT is transferable among various DNNs architectures in all input space categories. The results also demonstrate that when there is no attack, the performance of 1D-CNNs is better than SAEs in network traffic classification.

%% file: content/resultanalysis.tex
The results demonstrate that ANT is effective against all classifiers in various input space categories. However, its effectiveness is not the same in different experiments. The results indicate that the port numbers in the transport layer headers are very sensitive for PC-HP classifier, and also packet classifiers are more vulnerable to Start\_AdvPad attack than End\_AdvPad attack. These observations reveal that when the transport layer header is included in the input, the classifier selects port number as a good feature for network traffic classification. Also, the start of payloads is more sensitive to classifiers than the end of them. We think this is because the start of the payloads has a specific structure in most protocols, but the end of them is often filled by data, which can be different between the same class packets. Since AdvPad attack imposes bandwidth overhead to each packet, reducing the performance of packet classifiers requires more bandwidth overhead among other classifiers. However, the performance of PC-P is low, and PC-HP is vulnerable to port attack, which is a very simple attack and does not impose any bandwidth overhead to network traffic.

AdvPay attack imposes the least bandwidth overhead into network traffic among all attacks, and it significantly decreases the recall of all classes.  Hence, flow content classifiers have the lowest robustness among all classifiers. However, flow content classifiers are almost robust against the port attack. 
Since AdvBurst attack only injects some dummy packets with predetermined packet size or inter-arrival time into network traffic, the content of dummy packets can be anything. Therefore AdvBurst attack is the most challenging attack to detect or mitigate.

%% file: content/discussion.tex
In the image classification domain, humans determine the true class of image, and adversarial perturbation must be imperceptible to humans and does not change the true class of a sample. However, in network traffic classification, the true class of network traffic is determined by the application that generates and receives that network traffic. If the functionality of the application does not disrupt, the true class of network traffic is preserved. Proposed attacks change the content of network traffic to evade DL-based network traffic classifiers. However, since they do not change or remove the content of original network traffic, it can be retrieved. There are two approaches to retrieve original network traffic from perturbed network traffic. First, applications that generate and receive network traffic must be aware of the perturbation being added to network traffic and must remove this perturbation to retrieve original network traffic. Second, there must be a proxy to add or remove perturbation to network traffic. This proxy can be in the adversary's device or middleware in the adversary' network. The second approach is more convenient and is independent of applications that generate and receive network traffic.
Although some proposed attacks may not be applicable to any network situations, they have the potential to adapt to the various network situations by a little change.
For example, when a firewall in the adversary's network only passes the encrypted flows that start with SSL/TLS handshake packets, the adversarial pad can only be added to the end of the handshake packets, or dummy packet can be added after the handshake packets.

Previous studies \cite{DBLP:journals/soco/LotfollahiSZS20,DBLP:journals/access/MartinCSL17,DBLP:journals/access/WangYCQ18,DBLP:journals/cn/Caicedo-MunozEC18,DBLP:conf/icissp/Draper-GilLMG16,DBLP:journals/tnsm/AcetoCMP19,DBLP:journals/tifs/TaylorSCM18,8941027} have demonstrated that approaches being used to evade payload-based and port-based classifiers are not effective against the DL-based network traffic classifier. 
For example, although ISCXVPN2016 dataset consists of encrypted and tunneled network traffic, it is classified by the ML-based classifiers with high accuracy in this study an previous studies \cite{DBLP:journals/soco/LotfollahiSZS20,DBLP:journals/access/MartinCSL17,DBLP:journals/access/WangYCQ18,DBLP:journals/cn/Caicedo-MunozEC18,DBLP:conf/icissp/Draper-GilLMG16}. Similarly, in website fingerprinting attack, although the network traffic is generated by privacy-enhancing technologies such as Tor, it can be classified with high accuracy using time-series features of flows\cite{DBLP:conf/ndss/RimmerPJGJ18,DBLP:conf/ccs/SirinamIJW18}. 
Therefore proposed attacks have only focused on the evading of the DL-based network traffic classifiers, and if an adversary also wants to evade port-based or payload-based network traffic classifiers, (s)he must use appropriate approaches such as dynamic port assignment, payload encryption, or tunneling mechanism along with our proposed approaches to evade these classifiers. \\

%% file: content/conclusion.tex
In this paper, the robustness of DL-based network traffic classifiers against Adversarial Network Traffic (ANT) has been evaluated. 
Because of using universal adversarial perturbation generating methods in ANT, there is no need to have access to target network traffic in advance, and it is generated live. 
Based on the literature, we considered three input space categories in DL-based network traffic classification, including packet classification, flow content classification, and flow time series classification. Based on these categories, we proposed three new attacks to make ANT. The results indicate the robustness of network traffic classifiers in all input space categories are very low in facing ANT, and also, their performance considerably decreases by little random perturbation. 
The Robustness of DL-based network traffic classifiers is a critical issue, and by continuing researches in this area, we aim to increase the robustness of these classifiers.

%% file: content/Acknowledgement.tex
The authors would like to express their very great appreciation to Behrad Tajali, and Mohammad Reza Karimi for their valuable discussions, reviews, and feedback.